\documentclass[useAMS,usenatbib]{mnras}
\usepackage{graphicx}
\usepackage{epstopdf}
\usepackage{subcaption}
\usepackage{caption}
\usepackage{amsmath}
\usepackage{amssymb}
\usepackage{gensymb}
\usepackage[normal]{threeparttable}
\usepackage{calc}
\usepackage{pdflscape}
\usepackage{rotating}
\usepackage{afterpage}
\usepackage{color}
\title[WISE data as a photometric redshift indicator for radio AGN]{WISE data as a photometric redshift indicator for radio AGN}
\author[]{M. Glowacki$^{1,2,3}$\thanks{E-mail:
marcin@physics.usyd.edu.au}, J. R. Allison$^{1,4}$, E. M. Sadler$^{1,2}$, V. A. Moss$^{5,1,2}$ and T. H. Jarrett$^{6}$\\
$^{1}$Sydney Institute for Astronomy, School of Physics A28, University of Sydney, NSW 2006, Australia\\
$^{2}$ARC Centre of Excellence for All-sky Astrophysics (CAASTRO)\\
$^{3}$CSIRO Astronomy \& Space Science, PO Box 76, Epping NSW 1710, Australia\\
$^{4}$ARC Centre of Excellence for All Sky Astrophysics in 3 Dimensions (ASTRO 3D)\\
$^{5}$ASTRON, the Netherlands Institute for Radio Astronomy, Postbus 2, 7990 AA, Dwingeloo, The Netherlands\\
$^{6}$Department of Astronomy, University of Cape Town, Private Bag X3, Rondebosch, 7701, South Africa}
\begin{document}

%\date{Accepted YEAR MONTH DATE. Received ""; in original form ""}

\pagerange{\pageref{firstpage}--\pageref{lastpage}} \pubyear{2002}

\maketitle

\label{firstpage}

\begin{abstract}
We show that mid-infrared data from the all-sky WISE survey can be used as a robust photometric redshift indicator for powerful radio AGN, in the absence of other spectroscopic or multi-band photometric information. Our work is motivated by a desire to extend the well-known K-z relation for radio galaxies to the wavelength range covered by the all-sky WISE mid-infrared survey. Using the LARGESS radio spectroscopic sample as a training set, and the mid-infrared colour information to classify radio sources, we generate a set of redshift probability distributions for the hosts of high-excitation and low-excitation radio AGN. We test the method using spectroscopic data from several other radio AGN studies, and find good agreement between our WISE-based redshift estimates and published spectroscopic redshifts out to $z\sim1$ for galaxies and $z\sim3-4$ for radio-loud QSOs. Our chosen method is also compared against other classification methods and found to perform reliably. This technique is likely to be particularly useful in the analysis of upcoming large-area radio surveys with SKA pathfinder telescopes, and our code is publicly available. As a consistency check, we show that our WISE-based redshift estimates for sources in the 843\,MHz SUMSS survey reproduce the redshift distribution seen in the CENSORS study up to z~$\sim$~2. We also discuss two specific applications of our technique for current and upcoming radio surveys; an interpretation of large scale H{\sc i} absorption surveys, and a determination of whether low-frequency peaked spectrum sources lie at high redshift. 
\end{abstract}


\section{Introduction}

\begin{table*}
\centering
\caption{Examples of photometric redshift estimators for galaxies. Note that \citet{Bilicki2016} and \citet{DiPompeo2015} did not focus on radio galaxies. Many radio galaxies across the southern sky lack optical photometric redshift information, as well as K-band measurements at 2.2 $\mu$m.}
\label{tab:photoz}
\begin{tabular}{lcrr}
\hline
Study & z range & Method & Galaxy selection\\
\hline
\citet{Bilicki2016} & 0.002--0.7 & SuperCOSMOS, WISE and GAMA photometry & Optical\\
\citet{Burgess2006b} & 0.00183--2.852 & Optical r-band magnitudes & Radio-loud\\
\citet{DiPompeo2015} & 0.3--5.5 & SDSS \& WISE photometry & Optical\\
\citet{Donoso2009} & 0.4--0.8 & SDSS photometry via \citet{Collister2007} & Radio-loud\\
\citet{Willott2003} & 0.086--2.433 & K-band photometry & Radio-loud\\
\hline
\end{tabular}
\end{table*}

The field of astronomy is moving toward a new era of large radio surveys. These include (but are not limited to) the Evolutionary Map of the Universe \cite[EMU][]{Norris2011} to be conducted with the Australian Square Kilometre Pathfinder Telescope \cite[ASKAP;][]{Deboer2009, Johnston2009, Schinckel2012}, the Very Large Array (VLA) Sky Survey \cite[VLASS;][]{Murphy2015}, and the GaLactic and Extragalactic All-sky Murchison Widefield Array \cite[GLEAM][]{Hurley-Walker2017}. Most radio sources within such surveys are expected to lie at high redshift \cite[average redshift of z~$\sim$~0.8;][]{Condon1984}. 

While studies have been done on the local radio population, these are generally too shallow to capture the radio AGN population at high redshift or track their evolution with redshift. In contrast, deep but small area fields such as the Cosmological Evolution Survey \cite[COSMOS;][]{Scoville2007} are too small to contain significant numbers of the most luminous radio galaxies. Larger area surveys such as the Sloan Digital Sky Survey \cite[SDSS;][]{York2000}, Six-degree Field (6dF) Galaxy Survey \citep{Jones2009} and Galaxy And Mass Assembly survey \cite[GAMA;][]{Hopkins2013} are too shallow to capture the optical counterparts of radio populations that are more common at higher redshift. While a few exceptions exist such as the SDSS-Faint Images of the Radio Sky at Twenty-one centimeters (FIRST) sample \citep{Ivezic2002}, 2dF-SDSS Luminous Red Galaxy and QSO (2SLAQ) Luminous Red Galaxy Survey \citep{Sadler2007} and 2dF QSO Redshift Survey \cite[2QZ;][]{Boyle2000}, these are currently limited to their associated optical survey (in the first two instances SDSS, which only spans the northern hemisphere) and limited in redshift (0~$<$~z~$<$~2.35, 0~$<$~z~$<$~0.7, and 0.35~$<$~z~$<$~2.3 respectively).

Photometric redshift estimators are increasingly important, especially for radio AGN within the southern sky, many which lack any spectroscopic redshift information. Various photometric redshift studies exist (Table \ref{tab:photoz}) but some require optical information, such as that of \cite{Burgess2006} which used optical R-band magnitudes, and \cite{Donoso2009} which used SDSS photometry to make redshift estimates for radio sources hosted in early-type galaxies \cite[based upon the mega-redshift luminous red galaxies (MegaZ-LRG) photometric catalogue;][]{Collister2007}.

The K-z relation is another method, providing a clear correlation for radio AGN by using the K-band photometry at 2.2~$\mu$m \cite[e.g.][]{Lilly1982,Eales1997,Jarvis2001,DeBreuck2002,Willott2003,Inskip2010}. One limitation is that this work involved deep K-band observations, and existing K-band observations of radio galaxies south of declinations of 0$\degree$ are limited in sensitivity (e.g. the Two Micron All-Sky Survey - 2MASS). However, observations at similar wavelengths to the K-band can provide a relationship with the redshift of the radio source. For example, such a relation was found for infrared-faint radio galaxies at 3.6~$\mu$m \citep{Collier2014} with Spitzer.

The Widefield Infrared Survey Explorer \cite[WISE;][]{Wright2010} catalogue is an all-sky survey which can be used for photometric redshift estimation. A clear relation between the 3.4~$\mu$m WISE band and redshift has already been observed for radio-bright AGN \citep{Gurkan2014}, where a second-order polynomial fit was made to the narrow-line and low-excitation radio galaxies analysed in the study. Furthermore, \cite{DiPompeo2015} demonstrated the value of using WISE information in addition to SDSS photometry for quasars.

WISE goes deeper than other all-sky mid-infrared surveys \cite[e.g. compared to 2MASS, WISE will `go a magnitude deeper' or more for sources including galaxies,][]{Wright2010} and hence can probe the higher-redshift population. Furthermore, while not all radio sources are detected in WISE \cite[e.g. radio galaxies that are also faint in the mid-infrared,][]{Collier2014}, we can still place limits on the potential redshift of such radio sources, and hence provide useful information. Thus it is worth investigating the potential of the WISE information alone to assist future large radio surveys. 

In Section 2, we introduce the Large Area Radio Galaxy Evolution Spectroscopic Survey \cite[LARGESS;][]{Ching2017} sample which is employed as a training set for a method of estimating the redshift of any radio source with WISE data. This method involves class identification of the host radio galaxies and the calculation of the redshift probability distributions through a code we make publicly available\footnote{\url{https://github.com/marcinglowacki/wise_redshift_estimator}}. Section 3 examines the accuracy in estimating the redshift of the radio galaxy through our code using only the WISE information, through blind tests on spectroscopic samples with known classifications and redshifts. In Section 4 we present the results of two specific examples of applications the method offers; one on the SUMSS catalogue \citep{Mauch2003} in order to estimate the number and classes of radio sources that can be probed for associated H{\sc i} absorption, and another the sample of \cite{Callingham2017} to examine whether low-frequency peaked spectrum sources lie at high redshift. 

\begin{figure*}
\begin{subfigure}{1\textwidth}
\subcaption{(a) W1}
\includegraphics[width=1.0\linewidth]{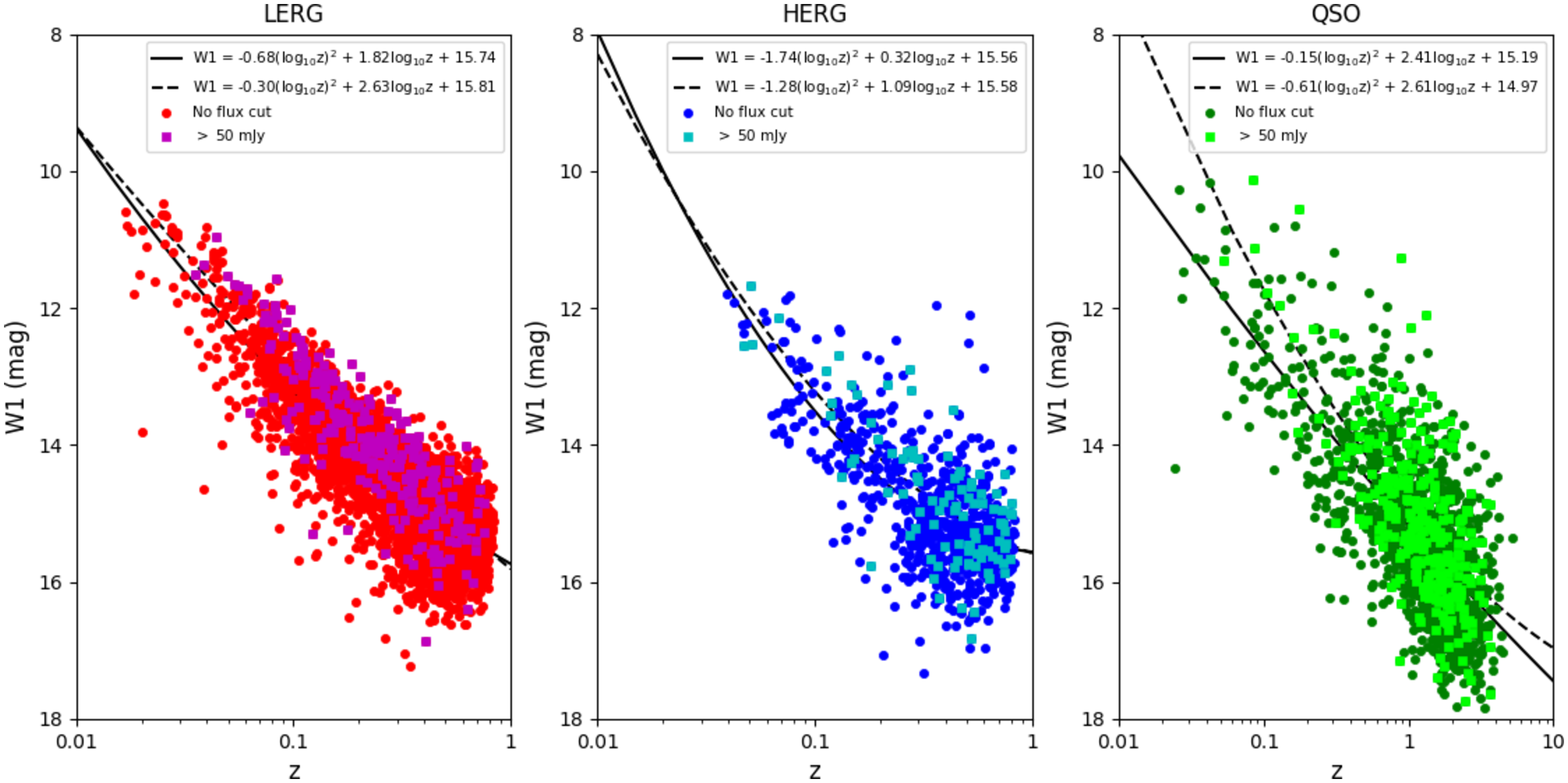}
\end{subfigure}
\begin{subfigure}{1\textwidth}
\subcaption{(b) W2}
\includegraphics[width=1.0\linewidth]{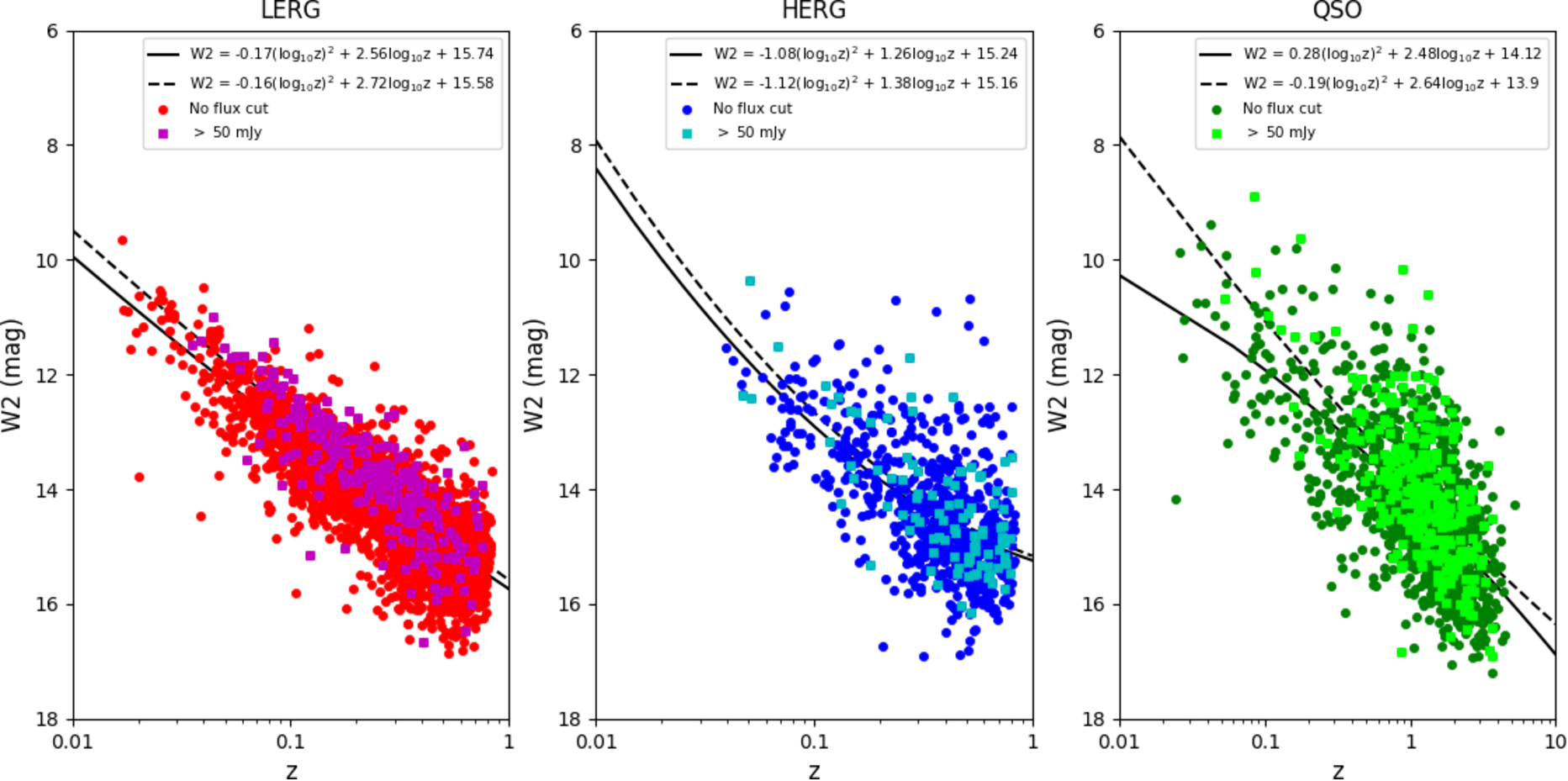}
\end{subfigure}
\caption{WISE magnitudes (W1 and W2) versus redshift for radio sources identified with low-excitation radio galaxies (LERGs), high-excitation radio galaxies (HERGs) and QSOs within the LARGESS sample. W3 and W4 relations are given on the following page. Note that here the LERGs and HERGs only extend up to z~$\sim$~0.8, while the QSO go up to z~$\sim$~5. The best fit equations to both all spectroscopically identified radio sources (solid line) and those with a flux density $>$ 50 mJy (dashed line) are given in each plot. Within the same class these fits are typically in good agreement, but differ significantly between different radio classifications in redshift evolution due to their different gas accretion methods. The clearest WISE-z relations with minimal flattening at higher redshift is seen for W1 and W2.}
\label{fig:WISE_z_fit}
\end{figure*}
\begin{figure*}
\ContinuedFloat
\centering
\begin{subfigure}{1\textwidth}
\subcaption{(c) W3}
\includegraphics[width=1.0\linewidth]{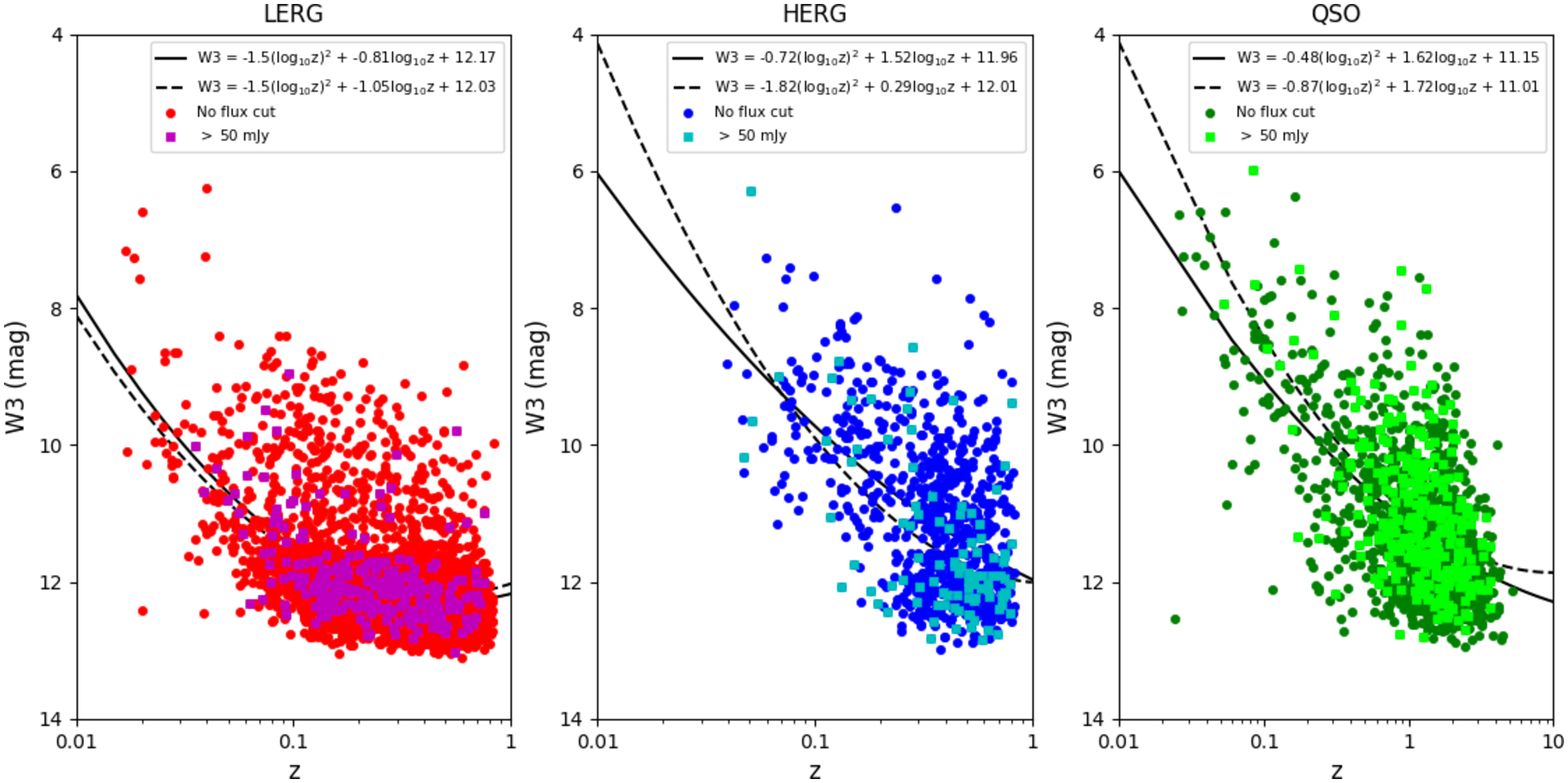}
\end{subfigure}
\begin{subfigure}{1\textwidth}
\subcaption{(d) W4}
\includegraphics[width=1.0\linewidth]{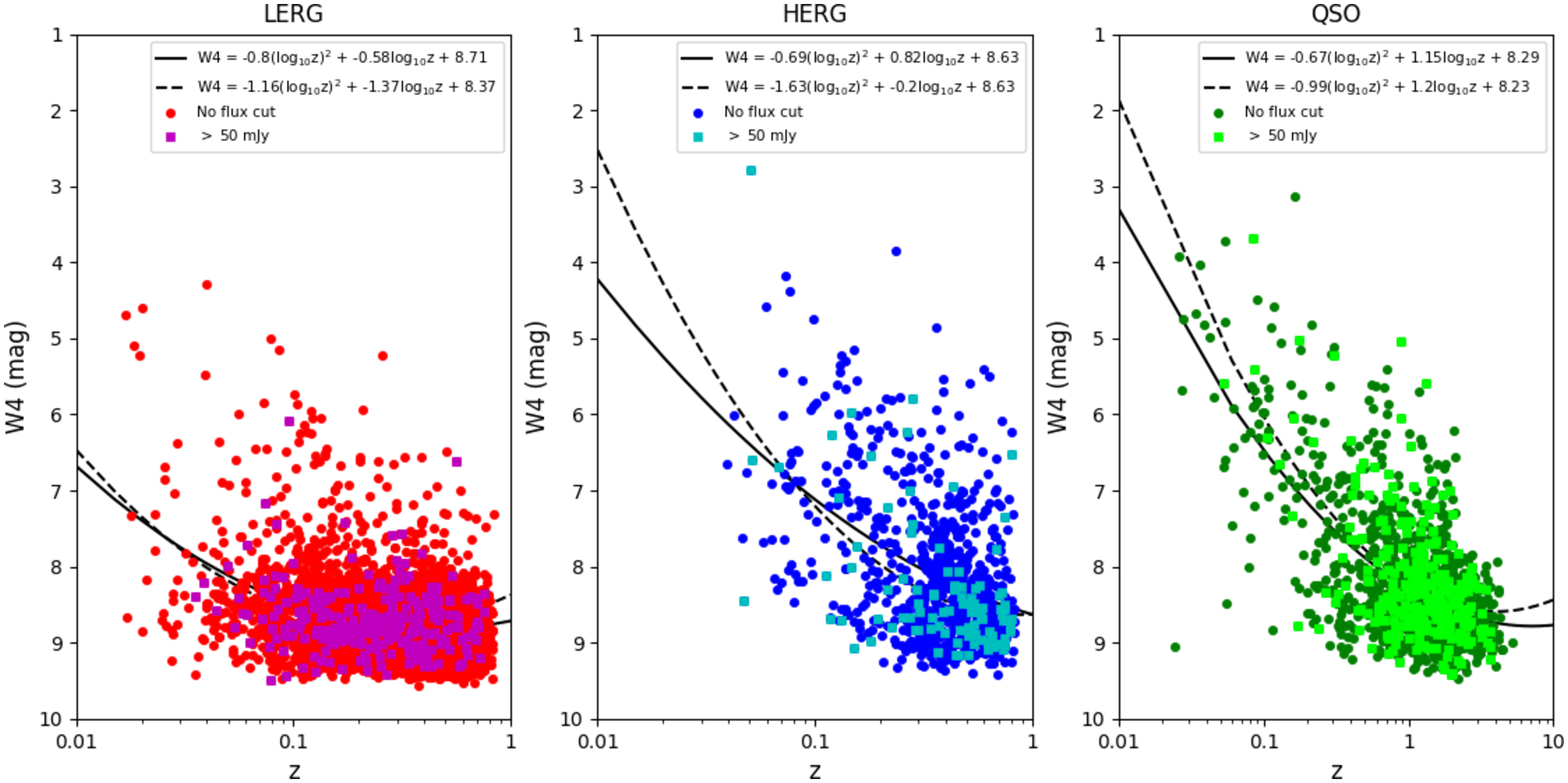}
\end{subfigure}
\caption{Continued. Relations for the W3 and W4 magnitude with redshift are given here. These relations are considerably flatter than those seen for W1 and W2 and are consistent with zero correlation (Section 2.1), and therefore are not useful for estimating the redshift of a radio source.}
\end{figure*}

\section{Method}

\begin{table*}
\centering
  \caption{The best-fit equations for W1 and W2 magnitudes versus redshift for the LERGs, HERGs and QSOs in the LARGESS sample. The top half gives W1 and W2 in terms of redshift, as done in \citet{Willott2003} and \citet{Gurkan2014}, and displayed in Fig.~\ref{fig:WISE_z_fit}. The bottom half gives the same relations as redshift in terms of W1 or W2. Differences can be seen across the different radio classifications. This provides a motivation to consider a redshift estimation method which accounts for the probability of a radio source being a LERG, HERG or QSO based on mid-infrared information.}
  \label{tab:bestfit}
  \begin{tabular}{lllc}
  \hline
Radio class & Best fit relation (W1) & Best fit relation (W2) & Fitted range\\
\hline
%LERG & \vbox{\begin{equation} -0.68(log_{10}~z)^{2} + 1.82log_{10}~z + 15.74 \end{equation}} & \vbox{\begin{equation} -0.17(log_{10}~z)^{2} + 2.56log_{10}~z + 15.74 \end{equation}}\\
LERG & W1 = --0.68\,(log$_{10}$\,$z$)$^{2}$ + 1.82\,log$_{10}$\,$z$ + 15.74 & W2 = --0.17\,(log$_{10}$\,$z$)$^{2}$ + 2.56\,log$_{10}$\,$z$ + 15.74 & 0 $<$ z $<$ 0.8\\
HERG & W1 = --1.74\,(log$_{10}$\,z)$^{2}$ + 0.32\,log$_{10}$\,z + 15.56 & W2 = --1.08\,(log$_{10}$\,$z$)$^{2}$ + 1.26\,log$_{10}$\,$z$ + 15.24 & 0 $<$ z $<$ 0.8\\
QSO & W1 = --0.15\,(log$_{10}$\,$z$)$^{2}$ + 2.41\,log$_{10}$\,$z$ + 15.19 & W2 = 0.28\,(log$_{10}$\,$z$)$^{2}$ + 2.48\,log$_{10}$\,$z$ + 14.12 & 0 $<$ z $<$ 4\\
\hline
LERG & log$_{10}$\,$z$ = 1.34 - 0.74(46.13 - 2.72\,{\rm W1})$^{0.5}$ & log$_{10}$\,$z$ = 7.53 - 2.94(17.26 - 0.68\,{\rm W1})$^{0.5}$ & 11~$<$~W1~$<$~15.5\\ 
%LERG ({\it EMS W1}) & log$_{10}$\,$z$ = $0.3337\,{\rm W1} - 5.3633$ & z = 0.000062$\times$10$^{0.26\,{\rm W1}}$ & 11~$<$~W1~$<$~15.5\\ 
%LERG (Marcin) & z = 0.000047$\times$10$^{0.26\,{\rm W1}}$ & z = 0.000062$\times$10$^{0.26\,{\rm W1}}$ & 11~$<$~W1~$<$~15.5\\ 
HERG & log$_{10}$\,$z$ = 0.09 - 0.29(108.40 - 6.96\,{\rm W1})$^{0.5}$ & log$_{10}$\,$z$ = 0.58 - 0.46(67.42 - 4.32\,{\rm W1})$^{0.5}$ & 12~$<$~W1~$<$~15.5\\
QSO & log$_{10}$\,$z$ = 8.03 - 3.33(14.92 - 0.60\,{\rm W1})$^{0.5}$ & log$_{10}$\,$z$ = --4.43 + 1.79(-9.66 + 1.12\,{\rm W1})$^{0.5}$ & 12~$<$~W1~$<$~16.5\\
\hline
\end{tabular}
\end{table*}

Recent studies have revealed evidence for two different populations of radio AGN \cite[see][and references within]{Best2012,Pracy2016}. Radiative radio AGN have efficient cold-mode accretion processes and are hosted in e.g. high-excitation radio galaxies (HERGs). These typically are associated with lower-mass galaxies and trace the evolution of star-forming galaxies. The radiatively inefficient mode of AGN, fuelled through the slow accretion of their hot halo gas, are hosted within low-excitation radio galaxies (LERGs) which typically trace the massive and passive elliptical galaxy population and found in rich environments compared to HERGs. These two populations were found to have separate evolution in luminosity and redshift \cite[e.g.][]{Pracy2016}.

Given the existence of these two populations with different luminosity and redshift evolution, an ideal photometric redshift estimator would be built on a survey which a) characterises radio galaxies in different radio classifications and addresses their separate redshift evolution, and b) has sufficient redshift coverage.

\begin{figure*}
\begin{subfigure}{1\textwidth}
\includegraphics[width=1.0\linewidth]{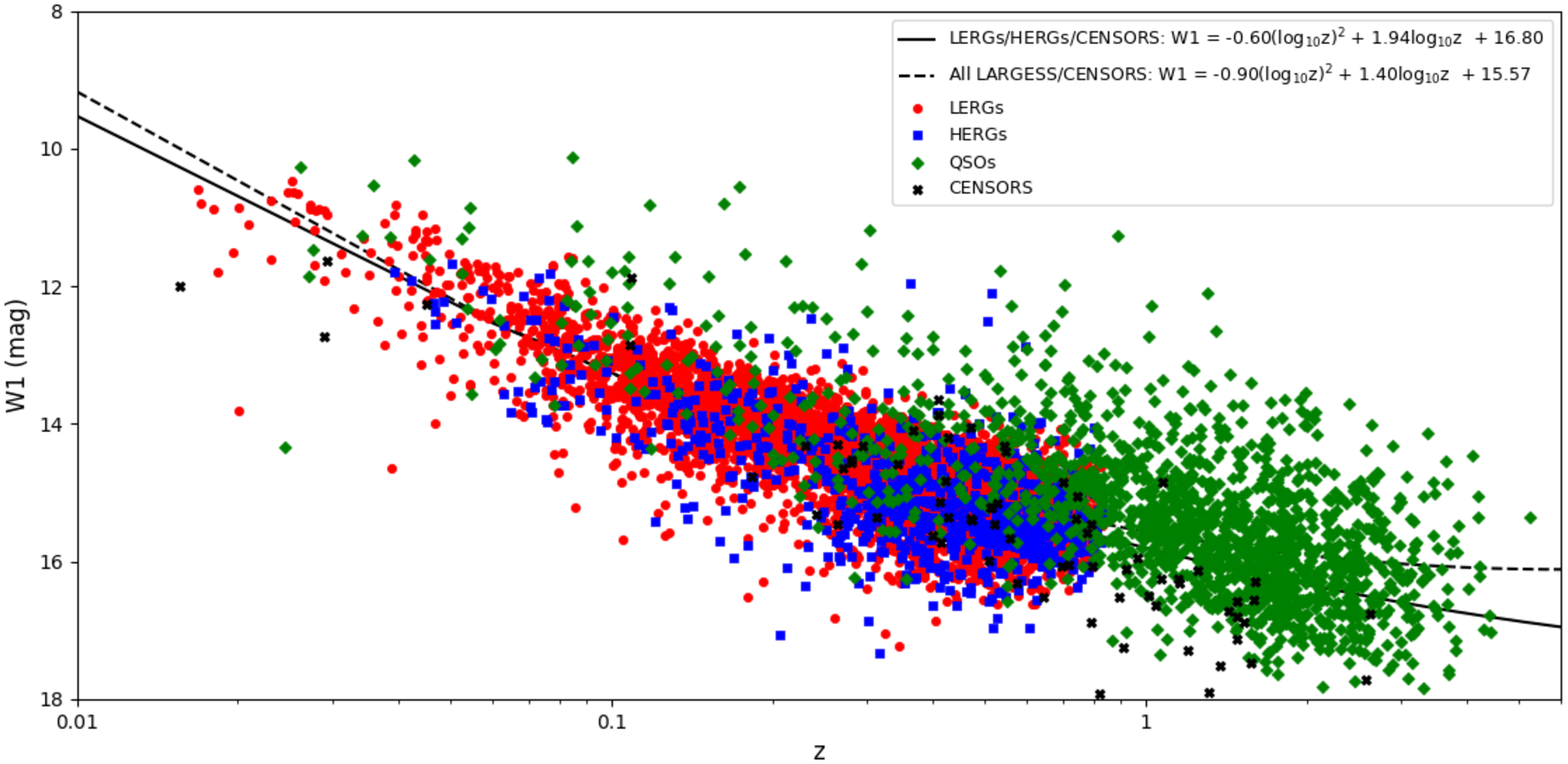}
\end{subfigure}
\begin{subfigure}{1\textwidth}
\includegraphics[width=1.0\linewidth]{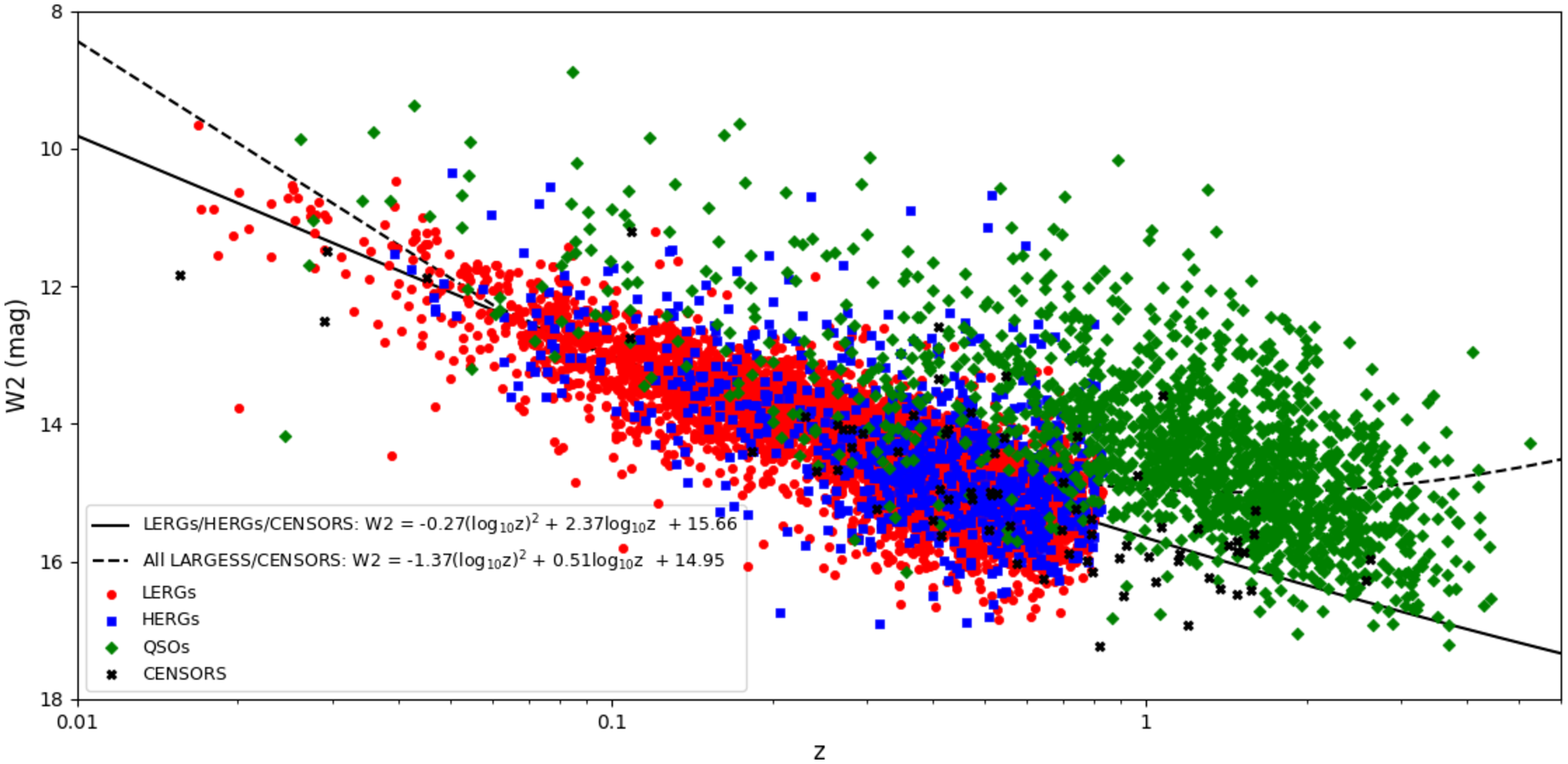}
\end{subfigure}
\caption{Quadratic fits to the LARGESS sample and CENSORS data for W1 and W2 magnitude versus redshift. We make two fits; one to the LERGs, HERGs and CENSORS points, and the other to all of LARGESS and CENSORS - that is, including the QSO population (Table~\ref{tab:loglinfits}). The CENSORS data extends the redshift range of fits made in Fig.~\ref{fig:WISE_z_fit} for the LERGs and HERGs, but it should be noted that redshift estimates are included here \citep{deZotti2010}.}
\label{fig:loglin_fit}
\end{figure*}

\begin{table}
\centering
  \caption{Principal component analysis (PCA) of the four WISE bands with redshift listed in order of significance (or decreasing eigenvalues, which sum to 5.0). Here we present the PCA results for all spectroscopically identified sources (LERGs, HERGs and QSOs) within the LARGESS sample. The first three principal components were considered significant here, with the remaining two only amounting to 4.1\% of the cumulative distribution (excluded from the table). We find that the most important contributions to these components, which are in bold font (those with the largest absolute value; here we assume absolute values greater than 0.5 to be significant) are redshift, W1 and W2.}
  \label{tab:PCA}
  \begin{tabular}{rrrrrr}
  \hline
Eigenvalue & Redshift & W1 & W2 & W3 & W4 \\
\hline
%All & 3.47 & -0.29 & -0.49 & -0.50 & -0.47 & -0.45 \\
%& 0.99 & -0.76 & -0.27 & -0.06 & 0.42  & 0.42  \\
%& 0.41 & -0.56 & 0.44  & 0.50  & -0.25 & -0.42 \\
2.91 & -0.26 & \textbf{-0.51} & \textbf{-0.54} & -0.45 & -0.42 \\
1.37 & \textbf{-0.64} & -0.36 & -0.13 & 0.47  & 0.48  \\
0.51 & \textbf{-0.70} & 0.35  & 0.44  & -0.12 & -0.43\\
\hline
\end{tabular}
\end{table}

\begin{table*}
\centering
  \caption{Best-fit equations for W1 and W2 magnitudes versus redshift as in Fig.~\ref{fig:loglin_fit} for the LARGESS and CENSORS surveys, and the same fits for redshift in terms of W1 or W2 magnitude. These fits give a basic indicator of the mid-infrared-redshift relation.}
  \label{tab:loglinfits}
  \begin{tabular}{llll}
  \hline
Sources & Best fit relation (W1) & Best fit relation (W2) & Fitted range \\
\hline
%LERG & \vbox{\begin{equation} -0.68(log_{10}~z)^{2} + 1.82log_{10}~z + 15.74 \end{equation}} & \vbox{\begin{equation} -0.17(log_{10}~z)^{2} + 2.56log_{10}~z + 15.74 \end{equation}}\\
LERG/HERG/ & W1 = --0.60\,(log$_{10}$\,$z$)$^{2}$ + 1.94\,log$_{10}$\,$z$ + 16.80 & W2 = --0.27\,(log$_{10}$\,$z$)$^{2}$ + 2.37\,log$_{10}$\,$z$ + 15.66 & $0~<~z~<~1.6$ \\
CENSORS\\
LARGESS/ & W1 = --0.90\,(log$_{10}$\,$z$)$^{2}$ + 1.40\,log$_{10}$\,$z$ + 15.57 & W2 = --1.37\,(log$_{10}$\,$z$)$^{2}$ + 0.51\,log$_{10}$\,$z$ + 14.95 & $0~<~z~<~4$\\
CENSORS\\
\hline
LERG/HERG/ & log$_{10}$\,$z$ = 1.62 - 0.83(44.08 - 2.40\,{\rm W1})$^{0.5}$ & log$_{10}$\,$z$ = 4.39 - 1.85(22.53 - 1.08\,{\rm W1})$^{0.5}$ & $11~<{\rm W1}~<~15.5$ \\
CENSORS\\
LARGESS/ & log$_{10}$\,$z$ = 0.78 - 0.56(58.01 - 3.60\,{\rm W1})$^{0.5}$ & log$_{10}$\,$z$ = 0.19 - 0.36(82.19 - 5.48\,{\rm W1})$^{0.5}$ & $11~<{\rm W1}~<~16.5$\\
CENSORS\\
\hline
\end{tabular}
\end{table*}

\begin{figure}
\begin{subfigure}{0.45\textwidth}
\subcaption{(a)}
\includegraphics[width=1.0\linewidth]{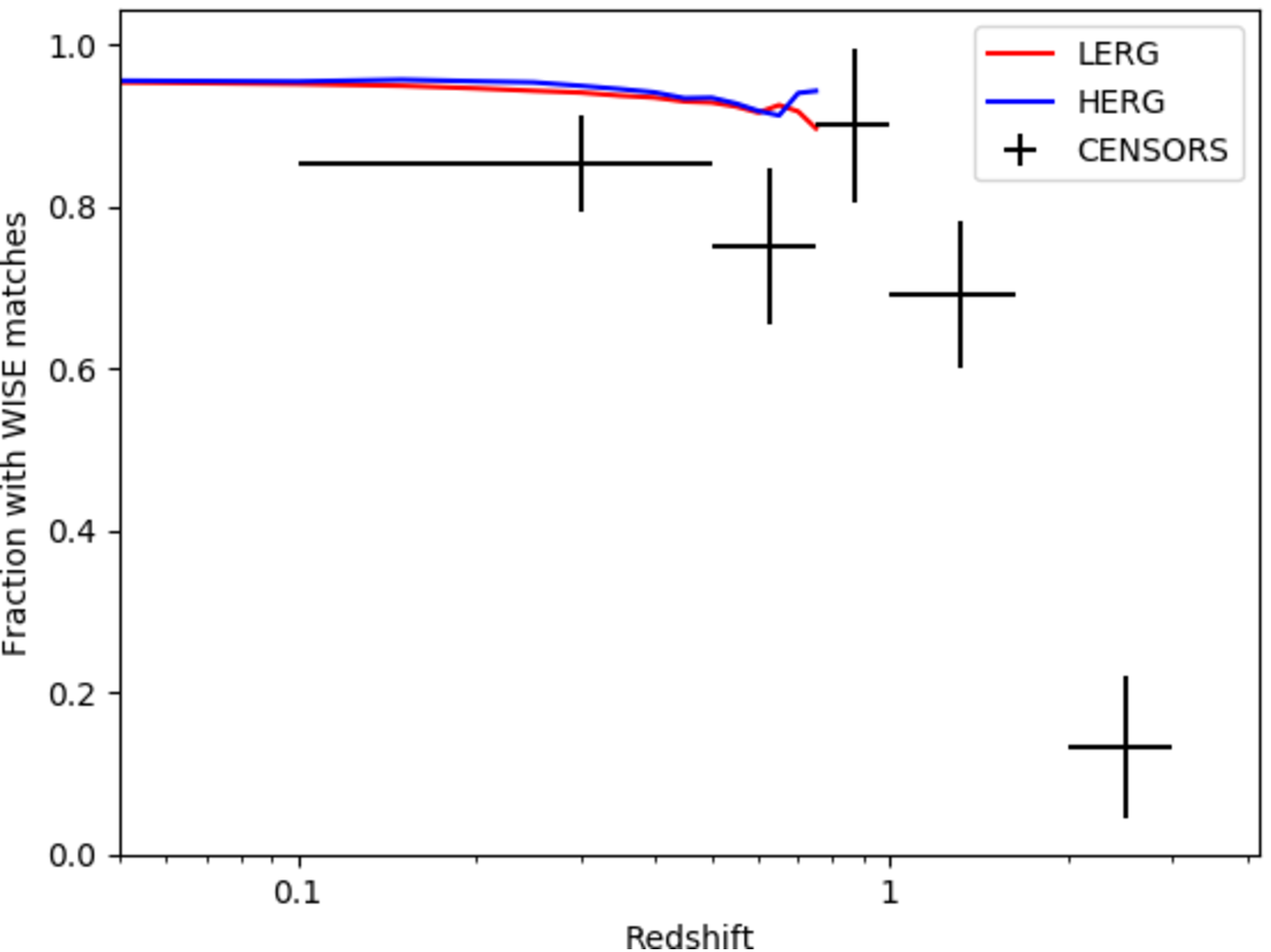}
\end{subfigure}
\begin{subfigure}{0.45\textwidth}
\includegraphics[width=1.0\linewidth]{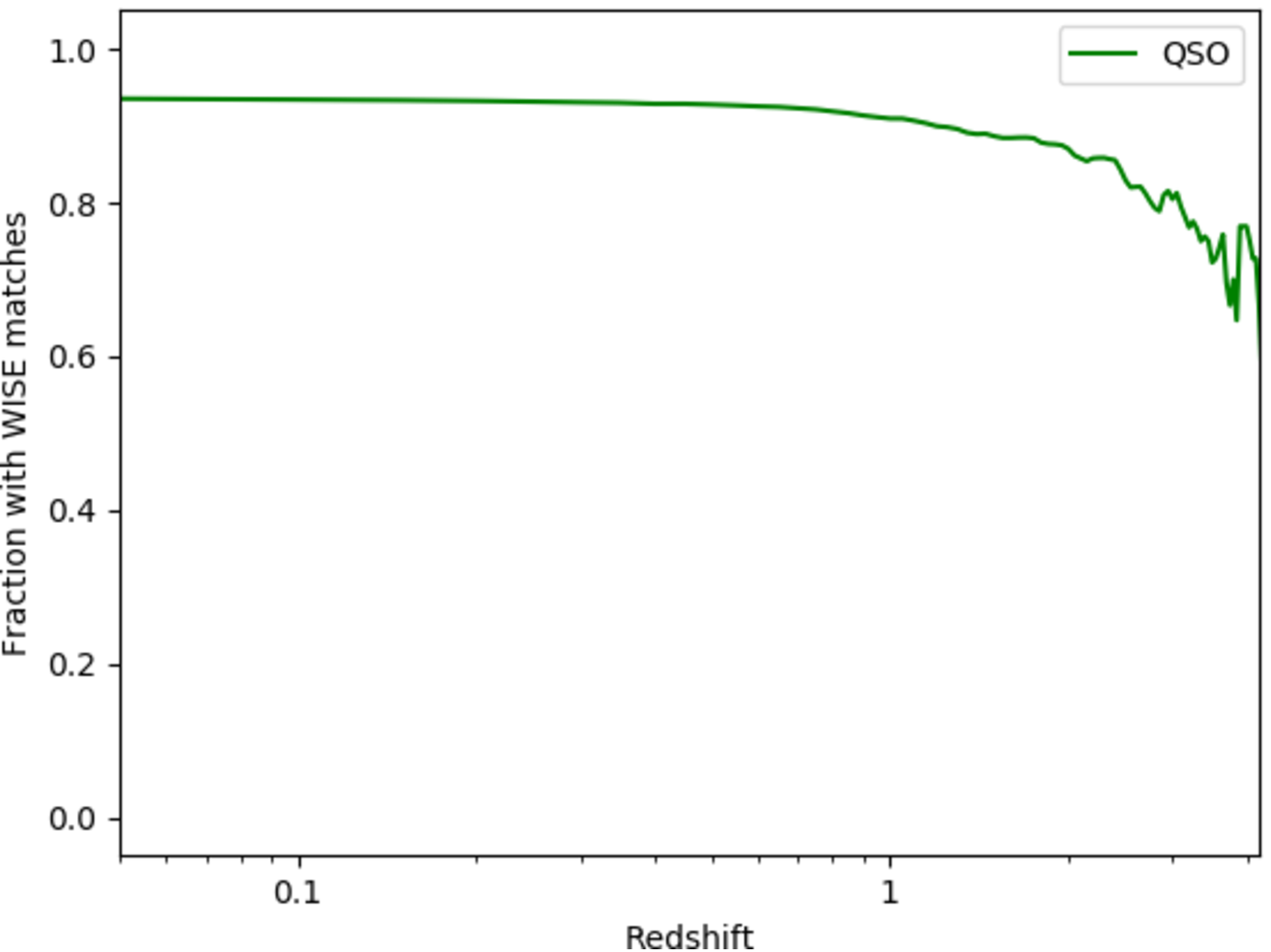}
\end{subfigure}
\begin{subfigure}{0.45\textwidth}
\subcaption{(b)}
\includegraphics[width=1.0\linewidth]{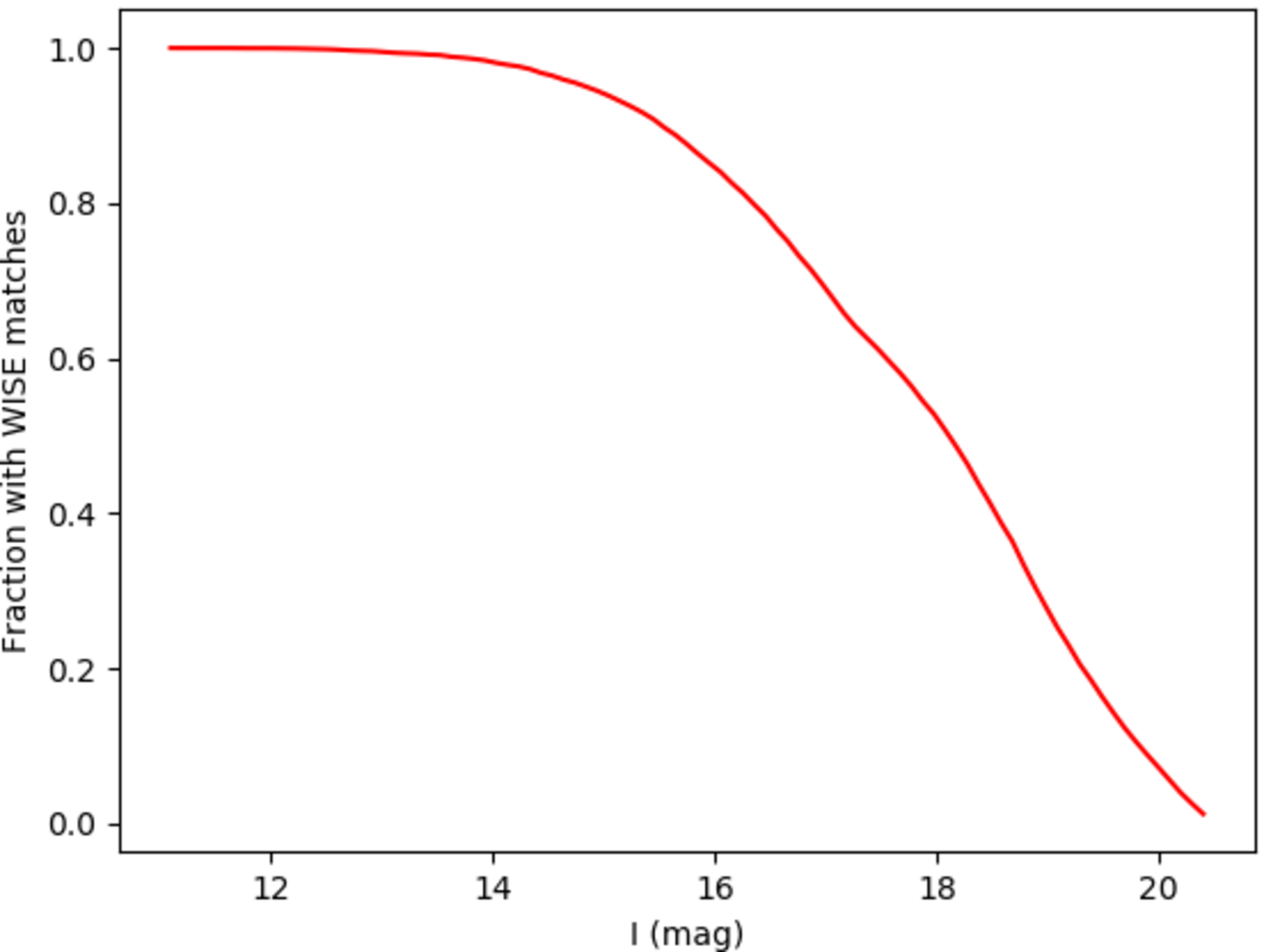}
\end{subfigure}
\caption{The fraction of sources with a WISE match within 3\,arcsec of LARGESS sources \citep{Ching2017} with a redshift measurement and LERG/HERG/QSO classification against a)  redshift (top and middle panels), and b) against the I-band magnitude (bottom panel). The WISE cross-match fraction is also given for the CENSORS survey in the top panel \citep{Brookes2008,deZotti2010}. The fraction is seen to steadily, albeit slowly, decrease with redshift. This suggests that the completion of the sample is good and not the cause of the upturn at higher redshift for fits made in Fig.~\ref{fig:WISE_z_fit}. Low number statistics is responsible for fluctuations seen in the HERG/LERG population at z~$>$~0.75, and the QSO population at z~$>$~4.}
\label{fig:complete}
\end{figure}

\begin{figure}
\includegraphics[width=1.0\linewidth]{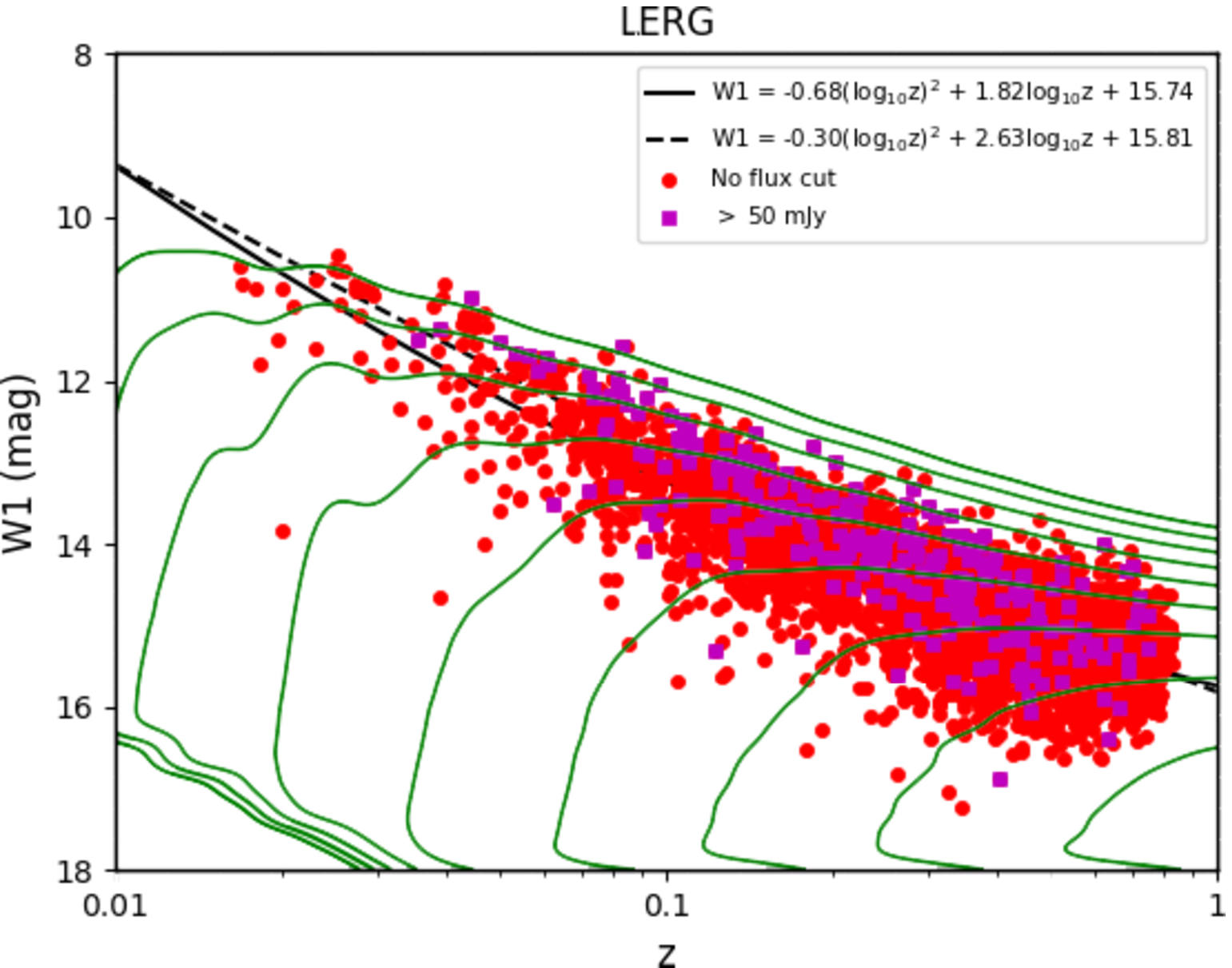}
\caption{Comparison of the W1 magnitudes versus redshift between the LARGESS LERGs and modelling done in fig. 9b of \citet{Jarrett2017}, represented with green contours. We find good agreement with the trends which suggests the slight upturn seen in many of the fits at high redshift (Fig.~\ref{fig:WISE_z_fit}) is real and not an effect of low-number statistics at z~$>$~0.7 for the LERG and HERG population (due to sensitivity limits).}
\label{fig:jarrett_overlay}
\end{figure}

The LARGESS sample \citep{Ching2017} was used as the training set for this work. LARGESS is a spectroscopic catalogue of radio sources designed to be representative of radio AGN populations out to at least redshift z~$\sim$~0.8. It contains optical identifications for 19,179 radio sources from the 1.4 GHz Faint Images of the Radio Sky at Twenty-cm \cite[FIRST;][]{Becker1995} survey down to an optical I-band magnitude limit of i$_{\rm mod}$~$<$~20.5 in Sloan Digital Sky Survey \cite[SDSS;][]{York2000} images, with no colour cuts. 

Furthermore, a WISE two-colour plot (W1-W2 versus W2-W3) can separate sources with different properties, e.g. those with recent star formation signatures (bright W3 magnitudes), or the presence of circumnuclear dusty regions heated by AGN activity (W1-W2~$>$~0.8). This results in clear distinctions between the LERGs (typically associated with passive elliptical galaxies), HERGs (associated with smaller but higher star-forming galaxies hosting radiative AGN) and QSOs in WISE colour space \cite[fig. 15 of][]{Ching2017}. Mid-infrared colour information unfortunately does not help with determining the redshift of the radio source, due to overlap in redshift evolution in mid-infrared colour space \cite[fig. 1 of][]{Blain2013}. 

LERGs comprise the majority (83\%) of LARGESS radio sources at z~$<$~0.8, with 12\% being HERGs and 5\% radio-loud quasi-stellar objects (QSOs) which extend to higher redshifts (z~$>$~5). The comparatively large fraction of LERGs agrees with previous studies within the same redshift range \cite[e.g.][]{Best2012}. Reliable spectroscopic redshifts and WISE data was crossmatched in \cite{Ching2017} for 9,294 sources in the sample (7,927 classified as a LERG, HERG or QSO), with the reliability of the WISE cross-matching estimated to be $\sim$99\%. Fig.~\ref{fig:WISE_z_fit} and Table~\ref{tab:bestfit} give the WISE magnitude versus redshift relations for each of the four bands.

\subsection{Determining the best WISE indicator of redshift}

The WISE survey has four observing bands: W1 (centred at 3.4 $\mu$m), W2 (4.6 $\mu$m), W3 (12 $\mu$m) and W4 (22 $\mu$m). We first investigate which of these four WISE bands would provide the best indicator of redshift. Given the already established K-z relation (2.2 $\mu$m), one could naively expect this relationship to hold for the WISE bands closest to this wavelength. The higher WISE bands are also less likely to provide a more robust relation with redshift than the bands at longer wavelengths \cite[e.g. W3 traces ISM heating, while W4 is the least sensitive band of WISE and mainly traces emission from warm dust;][]{Jarrett2017}. Nonetheless, we first investigated whether the other bands would be worth consideration.

We examine the Pearson correlation coefficients (PCC) between WISE magnitude and redshift, a measure of the linear correlation between two variables. Values of 1 (or -1) represent total positive (or negative) correlation, and 0 no linear correlation. We find PCC of 0.56 and 0.36 between the spectroscopic redshift and the W1 and W2 bands respectively for the LERGs, HERGs and QSOs in the LARGESS sample. The correlation was consistent with 0 for the W3 and W4 bands with redshift (correlation coefficients of -0.01 and 0.04). This is also evident in Fig.~\ref{fig:WISE_z_fit}, where the W3 and W4 bands offer a far flatter relationship with redshift. In all bands, a good agreement is generally found in he second order polynomial fits made between sources above 50 mJy in flux density at 1.4~GHz and all sources in the sample.

We performed a principal component analysis (PCA) between the four WISE bands and redshift to see which variables provide the strongest correlation with each other. Table~\ref{tab:PCA} gives the principal components for the LERGs, HERGs and QSOs in LARGESS. We find that while there is some correlation between redshift and all the WISE bands through PCA, the strongest correlations are seen between the two shorter-wavelength WISE bands and redshift. Hence we opted to use both W1 and W2 for our predictions of redshift to reduce computation time of our code. Redshift probability distributions (see the following subsection) generated from the W3 and W4 bands are also broader and offer less accuracy in redshift estimation. 

\subsubsection{Choice of redshift estimation method}

In order to compare between the three classes, we first make second-order polynomial fits to each of the WISE bands for each class in Fig.~\ref{fig:WISE_z_fit}, as done by \cite{Willott2003} (K-z relation) and \cite{Gurkan2014} (W1 versus z). The best-fit equations are given in the figure and Table~\ref{tab:bestfit}. However, these other studies only considered radio-loud AGN (e.g. through the 3CRR and 7CE samples), while with LARGESS we see that, within separate radio classifications, similar trends are found for both sources with a 1.4~GHz flux density~$>$~50~mJy and the whole sample (i.e. fainter radio sources). Furthermore, we see significantly different best fits (Table~\ref{tab:bestfit}) between the LERGs, HERGs and QSOs for each magnitude examined, and the relations do not match well for higher redshift. This highlights the differences between these populations, including in their redshift evolution \cite[e.g.][]{Best2012, Simpson2012, Best2014, Pracy2016}, which hence demands we treat the populations separately. For comparison, we also make fits to all LERGs, HERGs and QSOs in LARGESS alongside sources in CENSORS \citep{Brookes2008,deZotti2010} to extend the redshift fits for radio galaxies in Fig.~\ref{fig:loglin_fit} and give their relations in Table~\ref{tab:loglinfits}, by incorporating higher-redshift source information from the CENSORS survey \citep{Brookes2008,deZotti2010}.

A slight upturn at higher redshift is seen for the fits in the LERG and HERG populations for W1 and W2. This is believed to be a real effect of extragalactic WISE sources plateauing in W1 magnitude with higher redshift (e.g. host AGN at higher redshifts are brighter) \cite[see Fig.~\ref{fig:jarrett_overlay} which compares LARGESS with modelling by][in fig. 9b]{Jarrett2017}. This effect is supported over the hypothesis that this is a consequence of sensitivity limits in WISE for the LARGESS sample in Fig.~\ref{fig:complete}, which gives the fraction of LARGESS sources with redshift information and radio classification with a WISE cross-match against redshift, as well as the fraction of all LARGESS sources with a WISE cross-match in the full sample against I-band magnitude. There is only a small but steady decrease of the fraction of sources cross-matched with WISE with redshift. However, this slight effect could still bias the fits at higher redshift, and hence reduce the accuracy of any redshift estimation made through these fits.

Given these concerns on the best-fit relations found, rather than just offering a single best-fit equation, our method of redshift estimation takes advantage of the spectroscopic identifications done by \cite{Ching2017} and calculates the probability of a radio source to be a LERG, HERG and QSO. These probabilities are used to weigh the redshift estimation made for each class. 

As an alternative to the best-fit equation approach, we make available our code which creates a redshift probability density function across the full redshift range probed by LARGESS (see Section 2.3). By allowing for users to identify possible multiple peaks and observe the distribution widths, we offer more information on the potential redshift than a single number. This code requires only the W1, W2 and W3 magnitude information (W3 used only for radio classification; see following section) in order to generate a redshift probability density distribution, and can be found at \url{https://github.com/marcinglowacki/wise_redshift_estimator}. 

\subsection{Class identification}\label{kdeclass}

\begin{figure*}
\begin{subfigure}{0.49\textwidth}
\subcaption{(a) No flux cut}
\includegraphics[width=1.0\linewidth]{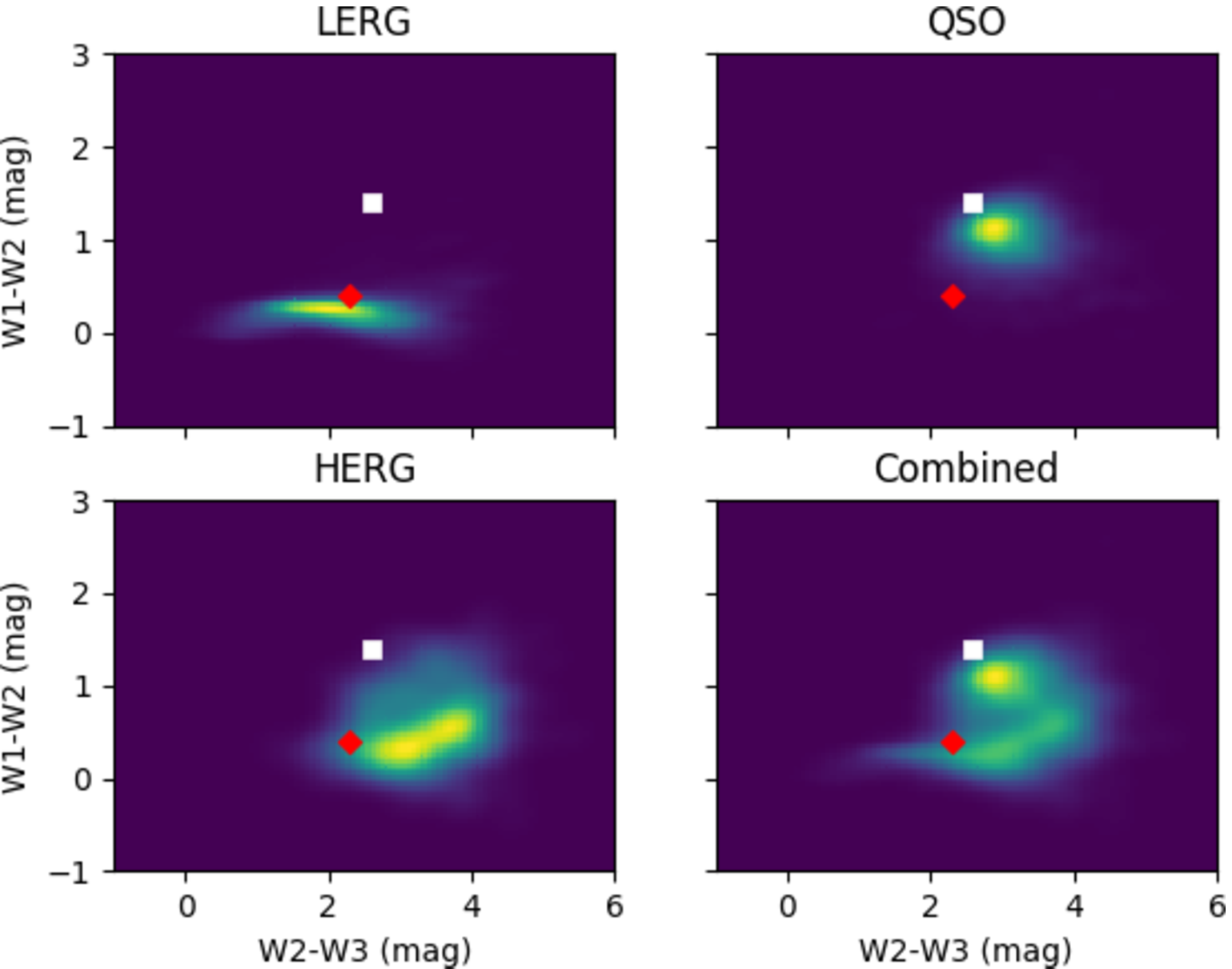}
\end{subfigure}
\begin{subfigure}{0.49\textwidth}
\subcaption{(b) Minimum 1.4 GHz flux of 50 mJy.}
\includegraphics[width=1.0\linewidth]{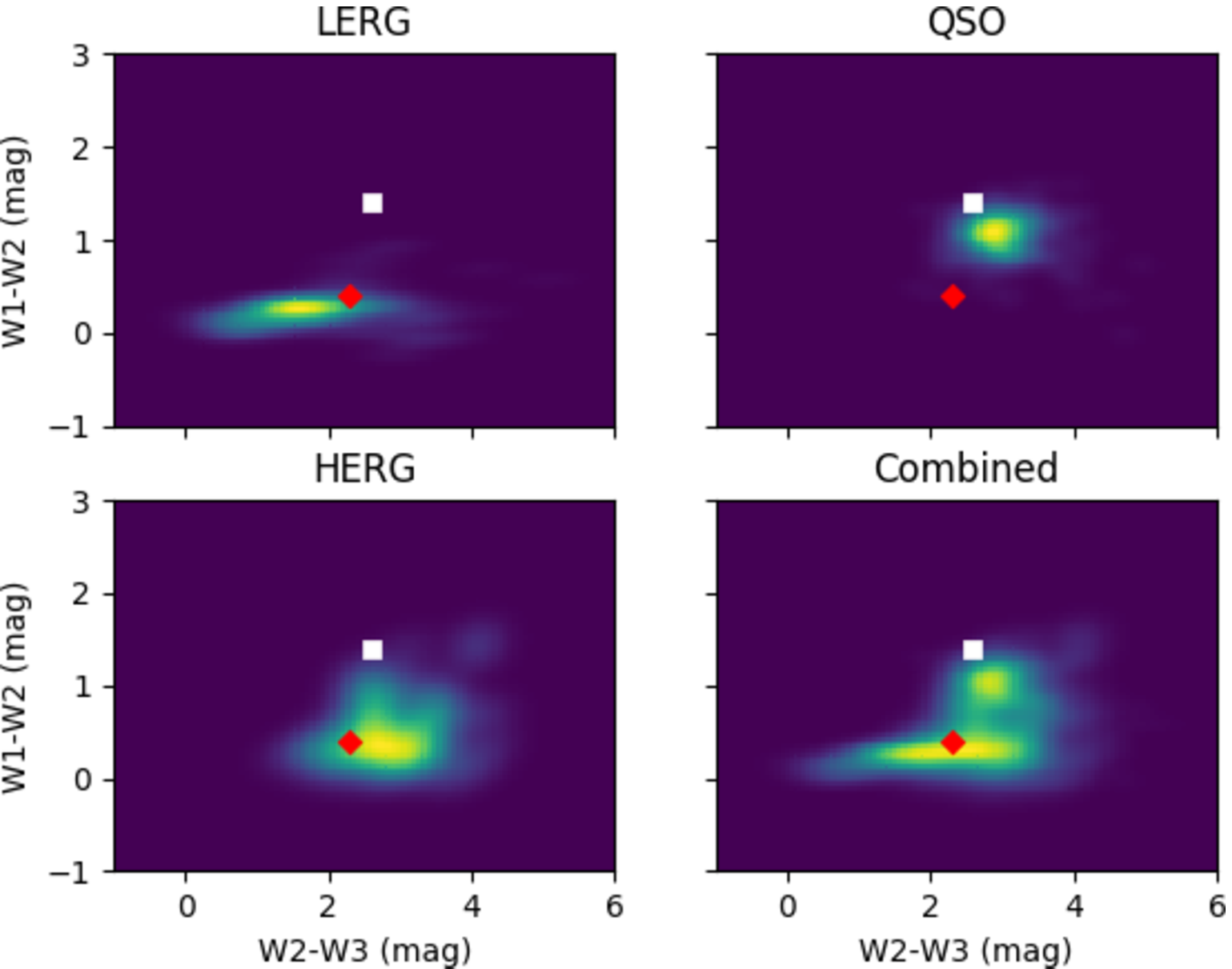}
\end{subfigure}
\caption{Distribution of WISE colours for the LARGESS sample. The left panel is for the LERGs, HERGs and QSOs in the full sample. The red diamond and white square represent the WISE colours for two test (hypothetical) radio sources with the same W1 measurement of 16~mag, referred to in Fig.~\ref{fig:probdist} as Case 1 and Case 2 respectively. As LERGs, HERGs and QSO sources have separate observed redshift evolutions, it is important to gain an indication of the radio classification. These objects occupy different regions of the WISE colour space. In this example the unknown radio source is likely to be either a LERG or HERG, rather than a QSO. The right panel shows sources in the LARGESS sample with an NVSS 1.4~GHz flux density of 50~mJy or greater. A minor shift to lower W2-W3 magnitudes is seen for the LERG and HERG populations, indicating less W3-bright galaxies remain in the sample from this limit.
}
\label{fig:wisecolours}
\end{figure*}

\begin{figure*}
\begin{minipage}{0.72\textwidth}
\centering
\captionsetup[subfigure]{aboveskip=-1pt}
\begin{subfigure}{1\textwidth}
\subcaption{(a) LERG}
\includegraphics[width=1.0\linewidth]{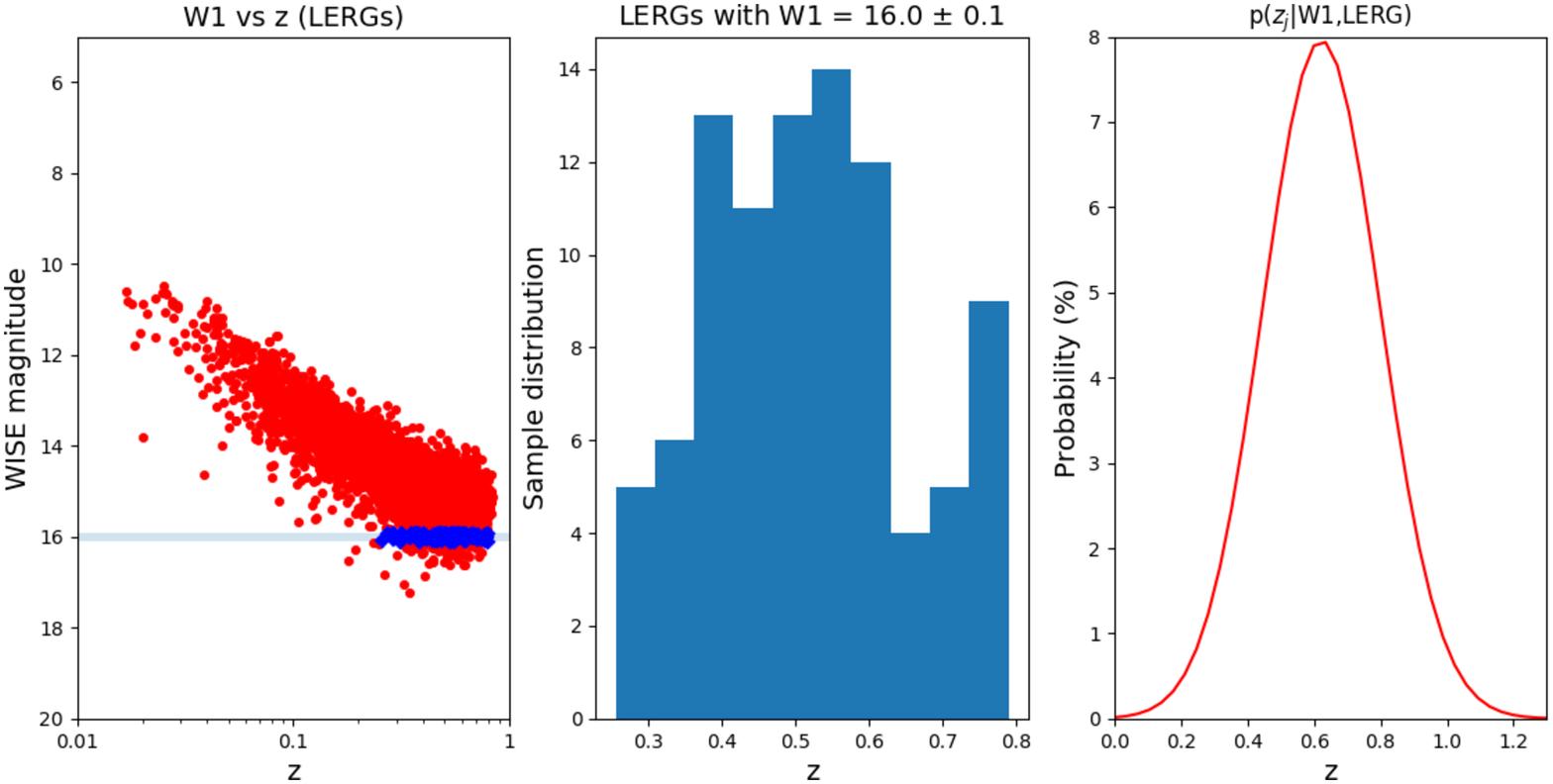}
\end{subfigure}
\begin{subfigure}{1\textwidth}
\subcaption{(b) HERG}
\includegraphics[width=1.0\linewidth]{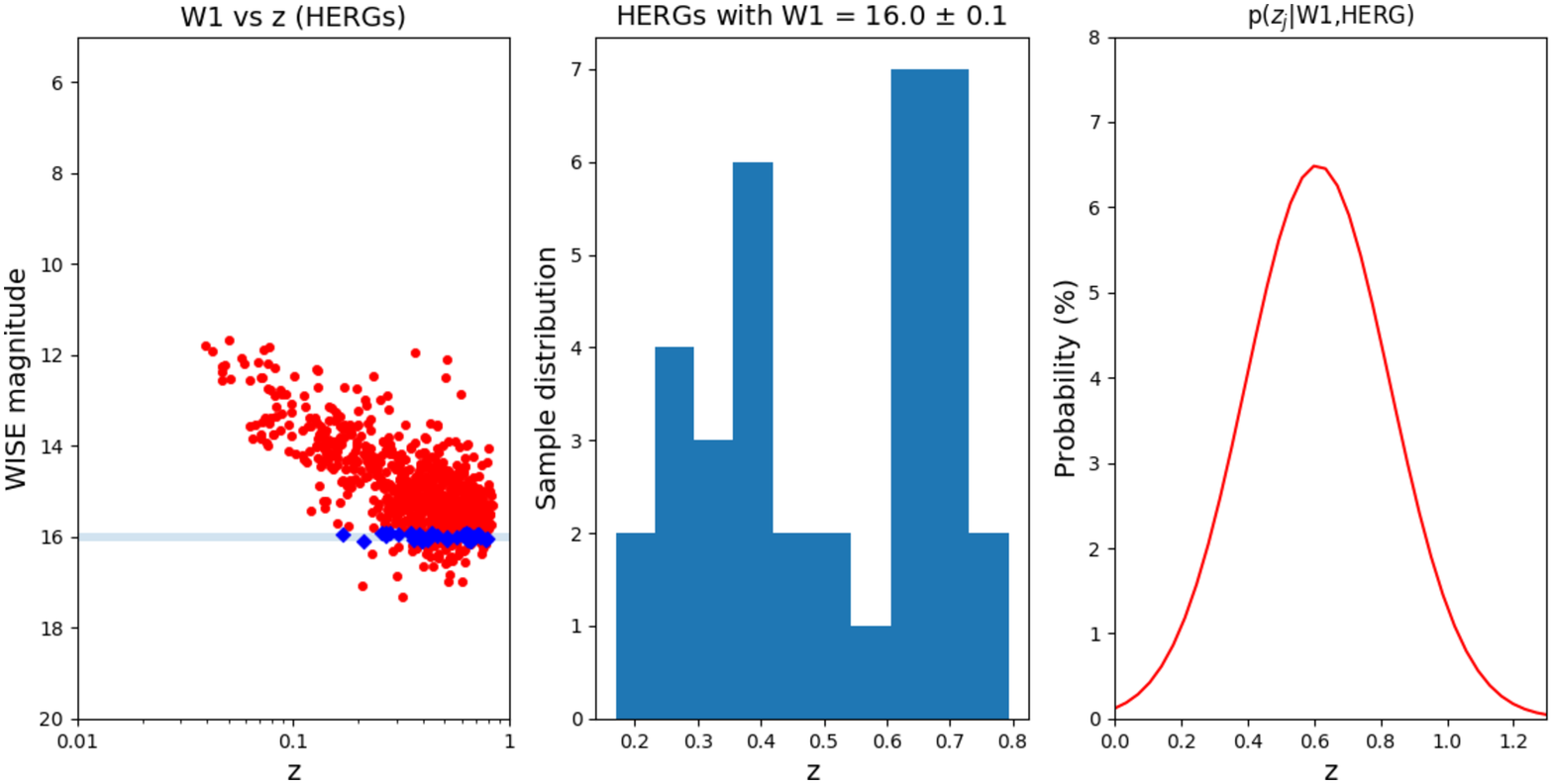}
\end{subfigure}
\begin{subfigure}{1\textwidth}
\subcaption{(c) QSO}
\includegraphics[width=1.0\linewidth]{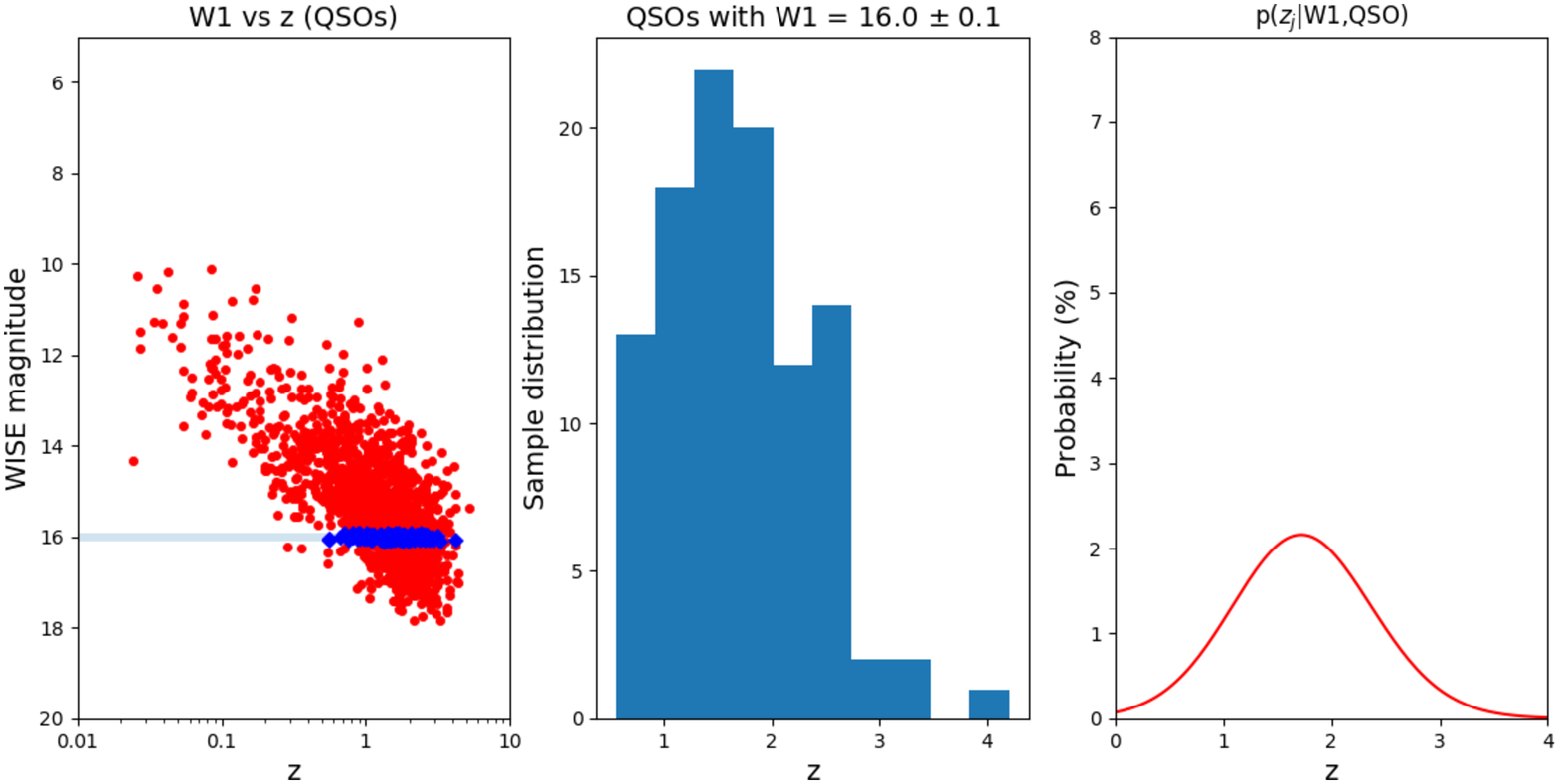}
\end{subfigure}
\caption{The distribution of the W1 WISE band with redshift for a hypothetical source, given the three possible class identifications of LERG, HERG and QSO. Left panel: The W1 magnitude versus redshift. Blue points are those within the W1 error magnitude of 0.1~mag of the test sources' W1 magnitude of 16.0~mag. Middle panel: Histogram representation of the blue points in the left panel. Right panel: The redshift probability distribution of the object, assuming it is of that class (i.e. not weighted by class probability), given the full LARGESS sample and weighted by the W1 magnitude. Note that in this example the distribution extends beyond z = 1 despite the limit of z = 0.8 in LARGESS for LERGs and HERGs, by extrapolating (Section 2.3.1). These probability distributions are also generated from the W2 magnitude information.}
\label{fig:distributions}
\end{minipage}
\end{figure*}

\begin{figure*}
\begin{minipage}{\textwidth}
\centering
\captionsetup[subfigure]{aboveskip=-1pt}
\begin{subfigure}{0.5\textwidth}
  \vskip 0pt
  \centering
\subcaption{(a) p(z), Case 1, W1}
\includegraphics[width=1\linewidth]{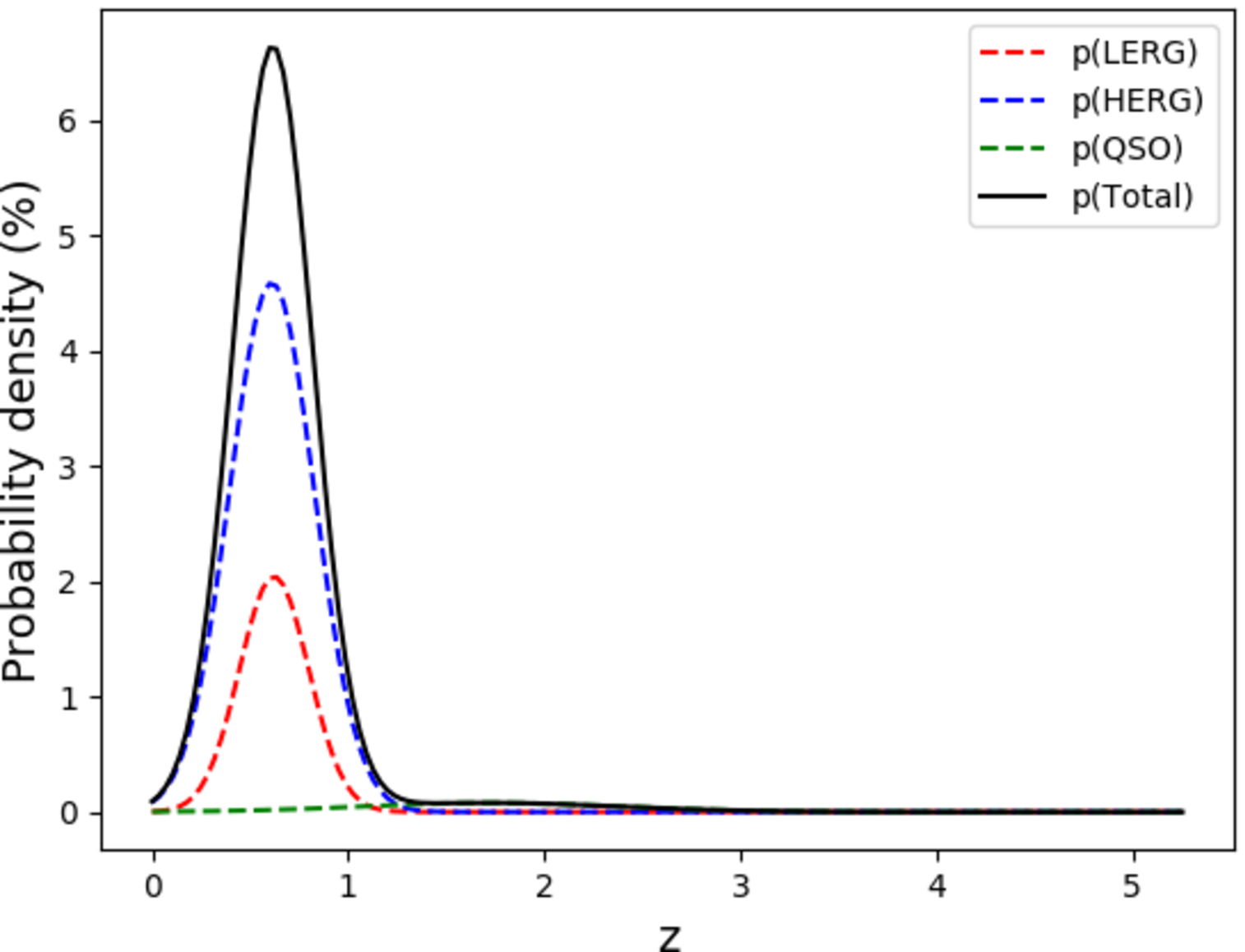}
\end{subfigure}%
\begin{subfigure}{0.5\textwidth}
  \vskip 0pt
  \centering
\subcaption{(c) p(z), Case 2, W1}
\includegraphics[width=1\linewidth]{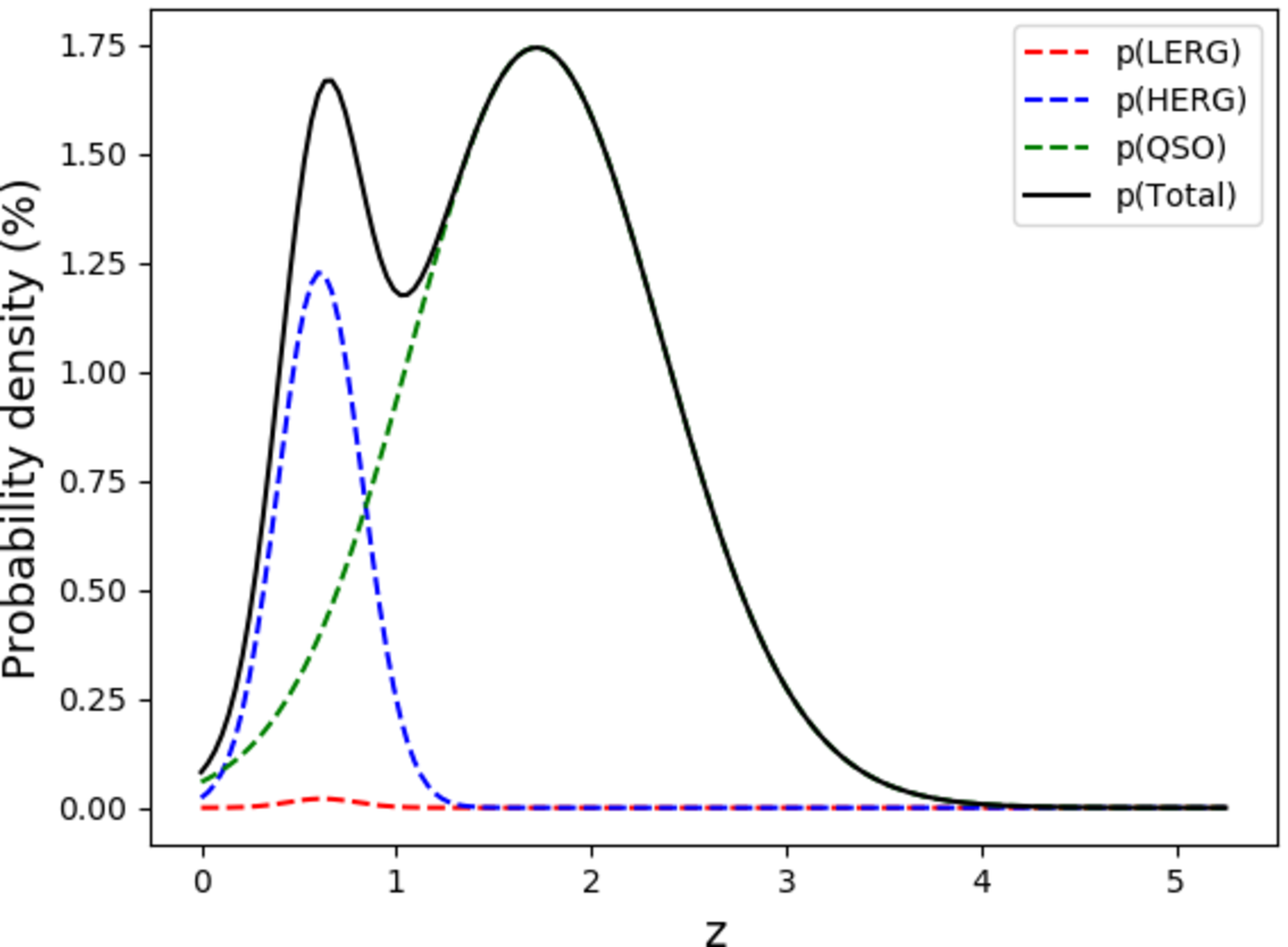}
\end{subfigure}
\begin{subfigure}{0.5\textwidth}
  \vskip 0pt
  \centering
\subcaption{(b) p(z), Case 1, W2}
\includegraphics[width=1\linewidth]{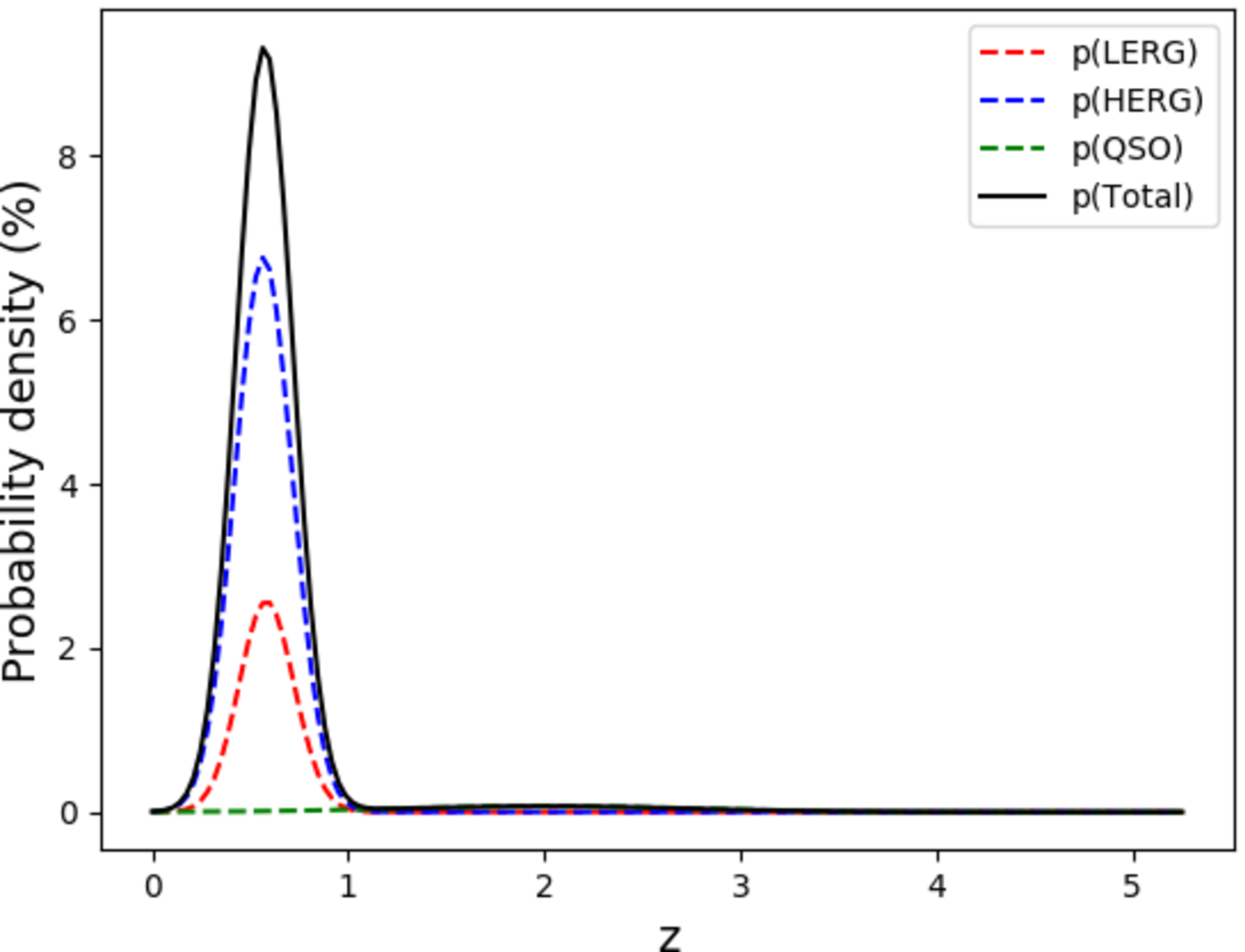}
\end{subfigure}%
\begin{subfigure}{0.5\textwidth}
  \vskip 0pt
  \centering
\subcaption{(d) p(z), Case 2, W2}
\includegraphics[width=1\linewidth]{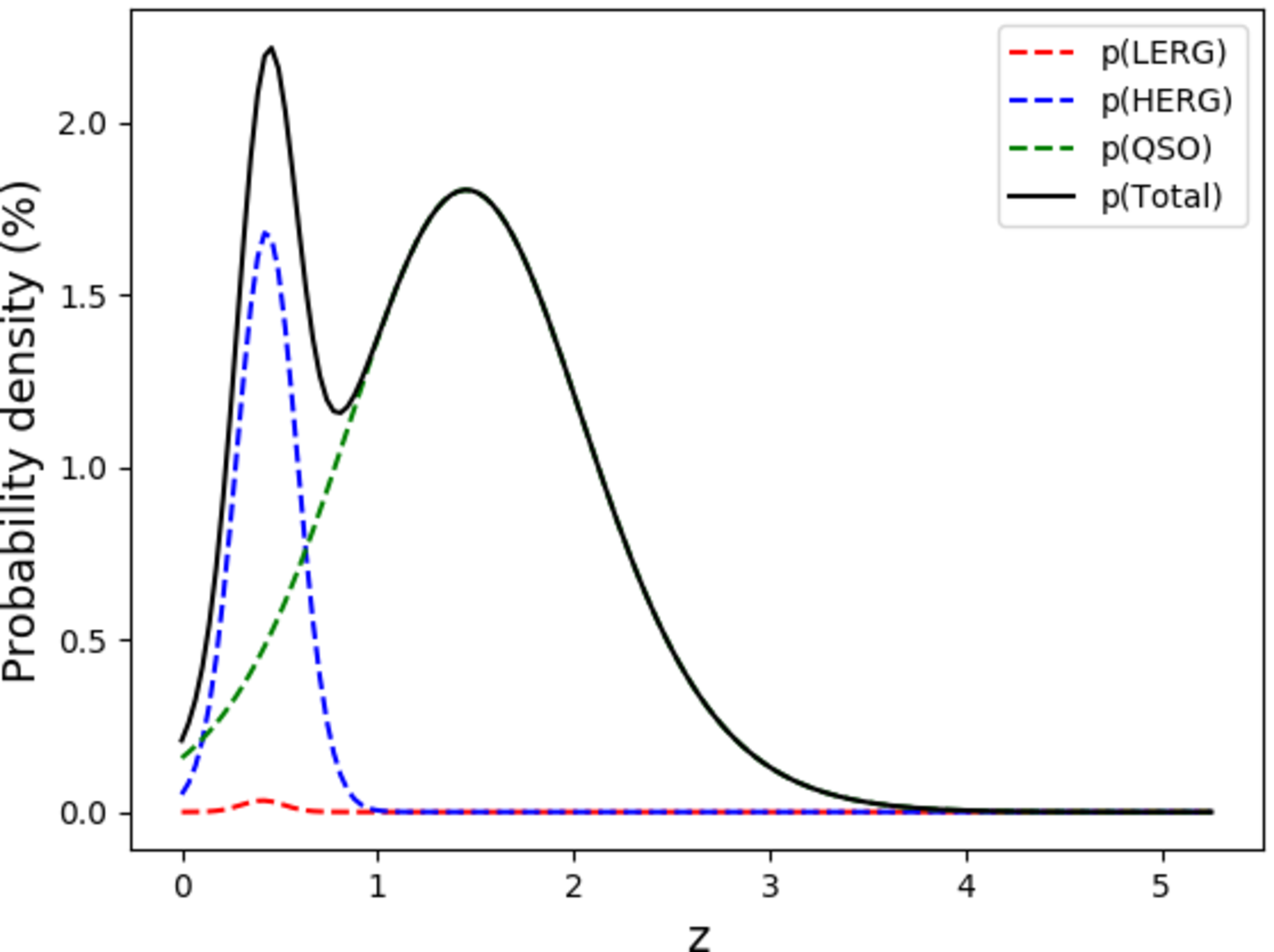}
\end{subfigure}
\caption{The redshift probability distribution for the example sources in Fig.~\ref{fig:wisecolours} and \ref{fig:distributions} for both W1 and W2 magnitudes (in Fig.~\ref{fig:wisecolours}, the radio source represented by the red diamond is Case 1, and white square Case 2). For Case 1, the higher likelihood of the object being a LERG or HERG rather than a QSO results in these two classes dominating the total probability curve. The class redshift probability densities (as seen in Fig.~\ref{fig:wisecolours}) are weighted by the probability determined from the WISE colour information. The distributions for Case 2 shows that the calculated probability of $\sim$81\% chance to be QSO from its WISE colour information dominates the resulting redshift probability distribution. There is good agreement between the two WISE bands.}
\label{fig:probdist}
\end{minipage}
\end{figure*}

In determining the probability for a radio source to be LERG, HERG or QSO through the WISE colour information, we use kernel density estimation (KDE). KDE is a non-parametric and smooth way to estimate the probability density function of random variables. It calculates the mean centre of input points and the distance for all points to this mean centre distributions. We use the KDE distributions in WISE colour-colour space to calculate probabilities for the radio sources (Fig.~\ref{fig:wisecolours}). The red diamond and white square represent example WISE colour points of two hypothetical objects. For each radio class, the probability of the object being said class is calculated depending on how far it lies from the KDE distribution peak, relative to the other two classes. From the KDE results, the source represented with the red diamond (Case 1) may be a LERG or HERG, but is unlikely to be a QSO. The source corresponding to the white square (Case 2) is most likely to be a QSO.

We note that the KDE method merely provides a probability to the object's class, which is influenced by outliers which occur within the LARGESS sample (e.g. AGN with unusual host galaxies). Furthermore, some radio sources have WISE colour information which place them away from any of these distributions, potentially due to unreliable measurements (e.g. sources with upper limits for their WISE magnitudes). In these cases we are unable to make a reliable estimation of class, and so the chance for the object being any of these three classes is assumed to be equal. This means that sources with unknown redshift and unreliable WISE W1, W2 and W3 magnitudes will lose accuracy in redshift estimation. We advise users to take care of this method for radio galaxies with unreliable WISE magnitude measurements. 

The distributions seen for the LARGESS sample are a combination of the true underlying distribution and measurement error. These errors would cause uncertainty in our redshift probability distribution, and possibly widen redshift confidence intervals. Deconvolution \cite[e.g. the `extreme' deconvolution - XDQSO - approach;][]{DiPompeo2015} would assist to minimise this effect; however, the errors on the WISE magnitudes in this training set are small. On average the error in WISE magnitude for the LARGESS sample is $<$~0.5\% for W1 and W2, and $\sim$1.5\% for W3. Therefore, uncertainties in the WISE magnitudes for the LARGESS sample would adversely affect the radio galaxy classification through the W2-W3 colour, rather than the redshift estimation directly (which uses W1 and W2 only). 

We note that radio flux density cuts can slightly change the WISE colour distribution; that is, the WISE colour and radio flux are correlated. The right panel of Fig.~\ref{fig:wisecolours} shows a flux limit of 50~mJy at 1.4~GHz (from FIRST and NVSS). While the QSO population does not change significantly, both the LERG and HERG populations shift to lower W2-W3 colours; ergo, to less dust-obscured or W3-bright galaxies, or the elliptical galaxy sector of the WISE colour diagram \cite[fig. 12 of][]{Wright2010}. This may be an effect of limiting to radio-bright sources which are typically associated with massive elliptical galaxies. It should also be noted that the source numbers are significantly lower here (670 objects compared to 7,927), which hence affects the reliability of redshift estimation. Our method makes use of the full LARGESS sample without flux cuts.

\subsection{Redshift probability density estimation}

We calculated a probability distribution in redshift for both the W1 and W2 bands, based on the LARGESS sample. In the case of the W1 magnitudes, for each class of radio AGN, we calculate the redshift probability distribution for each LARGESS source in W1 through a linear normal distribution fit (when appropriate; see Section 2.3.1 for cases with small number statistics). We then marginalise these redshift distributions over the range of magnitudes weighted by the observed W1 magnitude for each input source tested. The resulting distributions are weighted by the probability of the class as determined by the WISE mid-infrared colour information (that is, the chance for the object to be a LERG, HERG or QSO). For each source $i$, the redshift distribution is hence

\begin{equation}
p(z)_{i} = \sum_{C} \sum_{W1} p(z|W1,C) \cdot p(W1 | C)_{i} \cdot P(C)_{i},
\end{equation}

where $p(z|W1,C)$ is the redshift probability distribution for a given W1 value, marginalised over the probability of the given class $P(C)_{i}$. This is weighted by the observed W1 magnitude for the input source
\begin{equation}
p(W1 | C)_{i} = \exp (-\frac{(W1_{\rm val} - W1_{\rm mean})^{2}}{2 \cdot W1_{\rm std}^{2}}),
\end{equation}

where $W1_{\rm mean}$ is the W1 magnitude, $W1_{\rm std}$ the standard error in W1 for the source i, and $W1_{\rm val}$ the W1 value taken from the LARGESS sample. The same process is repeated for the W2 magnitudes. 

Fig~\ref{fig:distributions} gives the probability distributions calculated for individual classes for W1 for the same example sources as in Fig.~\ref{fig:wisecolours}, both with a W1 value of 16.0~mag and error of 0.1~mag. The left panel gives the W1 versus redshift relation for the LARGESS sample, with blue points falling within the range stated. The middle panel gives a histogram of these points for an input W1. W1 and W2 magnitudes fainter than 15.5~mag or brighter than 11~mag for all classes also fall in source numbers, and hence struggle to generate a proper redshift distribution for input objects with these WISE magnitudes. See the following subsection for details on extrapolation to account for these cases.

The three probability distributions are then multiplied by the class probabilities (Section 2.2) and marginalised to a final distribution (Fig.~\ref{fig:probdist}). Left panels are for Case 1 (red point in Fig.~\ref{fig:wisecolours}), and right panels for Case 2 (white point). Outputted information includes the median and the 68.26\%, 95.45\% and 99.73\% intervals (here-on written as 68\%/95\%/99.7\% intervals) of redshift for the source for both probability distributions. Calculation of the probabilities of the source having a redshift lower, higher and between values set by the user is also performed. 

Queries can be done for both individual or multiple objects through our code. Plots as seen in Fig.~\ref{fig:wisecolours} and \ref{fig:probdist} are optionally provided; these are more manageable for the user to consider for small sample sizes. Flux limits can also be placed on the LARGESS sample with the WISE colour determination step.

\subsubsection{Extrapolating for extreme W1 magnitudes}

As noted above, the LERG and HERG populations in LARGESS are limited to low redshifts due to sensitivity limits of the survey, with a clear upper limit of z~$\sim$~0.8 at W1 magnitudes fainter than 14 mag. Furthermore, the source counts in LARGESS drop off at W1 and W2~$<$~11 mag, meaning mid-infrared bright sources are unlikely to have a redshift distribution successfully generated. 

Given the visible trends in Fig.~\ref{fig:WISE_z_fit} and Fig.~\ref{fig:distributions}, it is fair to assume that objects with bright W1 or W2 magnitudes are likely to have redshifts of z~$<$~0.1, and those with WISE magnitudes fainter than 16 mag will have redshifts approaching and exceeding z~=~1, even for those assumed to be a HERG or LERG rather than a QSO. This is consistent with the deep source count modelling of the WISE extragalactic population in the GAMA G12 field \cite[see fig. 8 in][]{Jarrett2017}. In our method we extrapolate to faint WISE magnitudes and approximate the true distribution to higher redshifts for these populations.

We assume that the relationships between redshift and WISE magnitude in the first two bands (as well as the spread between z and WISE magnitude) in the well-sampled region are consistent and hold to higher redshift (Fig.~\ref{fig:WISE_z_fit}). These regions are 11~$<$~W1~$<$~15.5 for LERGs, and 13~$<$ W1~$<$~15.5 for HERGs (due to their significantly smaller population in the LARGESS sample), and likewise for W2. The trends are used to generate the redshift probability distribution for bright or faint WISE magnitudes. The remaining application of the method remains the same. Section 3.2.1 compares the accuracy and success in generating a redshift distribution when we extend to a larger redshift range to when we do not extrapolate. 

In the following section, we consider the accuracy of our method on different surveys with optical spectroscopic information, and examine how the method performs across different redshift ranges.

\section{Method verification}

\subsection{Test Samples}

In order to verify the accuracy of our redshift estimator algorithms, we conducted blind tests of our code on separate samples with known spectroscopic redshift information. The test samples are as follows (and summarised in Table~\ref{tab:testsurveys}):

\begin{table}
\centering
\caption{Spectroscopic surveys in which we compare the available redshift information for these sources with our predictions. Size is the number of sources in the sample; not all sources had a good cross-match with WISE with good quality measurements. We compare the CENSORS survey with redshift predictions for the SUMSS survey in Section 3.4.}
\label{tab:testsurveys}
\begin{tabular}{lrrc}
\hline
Sample & Size & Redshift range & Section\\
\hline
AT20G/6dF & 2,236 & 0.00~$<$~z~$<$~4.63 & 3.2, 3.3\\
Best \& Heckman & 9,136 & 0.01~$<$~z~$<$~0.30 & 3.2, 3.3\\
SDSS-DR12 Quasar & 2,761 & 0.04~$<$~z~$<$~5.25 & 3.3\\
CENSORS & 150 & 0.02~$<$~z~$<$~3.43 & 3.4\\
\hline
\end{tabular}
\end{table}

\begin{itemize}
\item \textbf{AT20G/6dF sample}: A sample of 2,236 radio sources selected from Australian Telescope 20 GHz (AT20G) sources followed up with optical spectroscopy \citep{Mahony2011} and the 6dF Galaxy Survey \citep{Jones2009}. The AT20G sources were primarily classified as QSOs (making up 40.9\% of the total test sample), while the 6dF sources were primarily LERGs (33.5\%) and some HERGs (5.66\%). The remaining sources were classified as either star-forming or just as galaxies, and so for these sources only the accuracy of the redshift prediction was examined. These sources have a redshift range of 0~$<$~z~$<$~4.63.
\item \textbf{Best \& Heckman sample}: The main sample of 9,136 radio AGN studied by \cite{Best2012} constructed through combining SDSS with FIRST. These included 6,047 LERGs and 216 HERGs, with the remaining sources not spectroscopically identified as either source in the study. This sample was at relatively lower redshifts, spanning 0.01~$<$~z~$<$~0.3.
\item \textbf{SDSS-DR12 Quasar sample}: FIRST radio sources cross-matched with the SDSS-DR12 quasar catalogue \citep{Paris2017}. This provided a sample of 2,761 visually inspected QSOs with spectroscopic redshifts with sufficient WISE magnitudes for our code. These sources were found in the redshift range 0.04~$<$~z~$<$~5.25, and were not spectroscopically separated into LERGs, HERGs and QSOs as in LARGESS.
\item \textbf{CENSORS sample}: The Combined EIS-NVSS Survey Of Radio Sources \cite[CENSORS;][]{Brookes2008} is a 1.4~GHz radio survey selected from the NRAO VLA Sky Survey \cite[NVSS;][]{Condon1998}, which is complete to 7.2~mJy and overlaps with the ESO Imaging Survey (EIS) Patch D. This sample and its redshift distribution \citep{deZotti2010} is compared with redshift predictions made for the Sydney University Molonglo Sky Survey \cite[SUMSS;][]{Mauch2003} in Section 3.4.
\end{itemize}

Each catalogue had their radio position cross-matched with WISE, although this was done already within the publicly available SDSS-DR12 quasar sample. We applied the same restrictions on quality and cross-match radius with this sample. Two cuts were placed on each sample following this cross-match. The first was placed on the separation distance between the radio and WISE source. An upper limit of 2.5\,arcsec was placed, in order to ensure higher quality results at the cost of completeness (see Section 3.4). 

The second cut was applied to sources with poor quality WISE magnitudes from the WISE catalogue. This aligns with that used for the LARGESS sample, which required a signal-to-noise ratio of at least 2 for all sources with WISE counterparts \citep{Ching2017}. We point out that sources with lower quality W1, W2 and/or W3 measurements typically result in a lower success rate for our redshift prediction (e.g. for the AT20G/6dF sample, success rates to contain the known spectroscopic redshift within the 68\% confidence interval were 60.1\% versus 61.9\% and 66.2\% versus 67.7\%, for W1 and W2 respectively). We hence emphasise the need to consider the quality of the WISE information of input objects, as low-quality or upper limit values on the magnitudes affect both the ability to accurately classify objects by their WISE colours and hence quantify their likely redshifts. Radio sources with upper limits in their W1 or W2 bands are also likely to have their redshift underestimated.

\subsection{Mid-infrared colour class identification}

\begin{table}
\centering
\caption{Radio classification success rates for the blind tests of the Best \& Heckman (2012) sample and the AT20G/6dF sample, both which had spectroscopic identifications for LERGs and HERGs (QSOs only in the latter sample). Overall the accuracy of the classification system is high. HERGs are the least successful to be classified, which is attributed in part to the smaller population of HERGs within the LARGESS training set, and its overlap in WISE colour space with LERGs and QSOs (Fig.~\ref{fig:wisecolours}).}
\label{tab:classrates}
\begin{tabular}{lrrrr}
\hline
Sample & LERG & HERG & QSO & Total\\
\hline
Best \& Heckman sample & 90.7\% & 60.9\% & - & 89.1\%\\
AT20G/6dF sample & 93.5\% & 50.9\% & 86.7\% & 87.0\%\\
\hline
\end{tabular}
\end{table}

\begin{table}
\centering
\caption{Redshift estimation success rates for the three test samples. The spectroscopically determined redshift was compared to the 68\%, 95\% and 99.7\% redshift confidence intervals generated through our method. Probability distributions in redshift predicted by the W2 magnitude had wider confidence intervals on average and hence has a higher success rate in most instances. The W1 and W2 confidence intervals have good agreement. These percentages do not match the confidence interval range as some radio sources may not be properly classified by their WISE colour information, thus affecting the redshift estimation accuracy, while in the Best \& Heckman sample the relatively low redshift range of the sample (z~$<$~0.3) and high classification rate resulted in higher success rates.}
\label{tab:redshiftsucc}
\begin{tabular}{lrrr}
\hline
WISE magnitude & 68\% & 95\% & 99.7\%\\
\hline
Best \& Heckman\\
W1             & 74.0\% & 97.2\% & 99.8\%\\
W2             & 77.6\% & 99.1\% & 99.9\%\\
\hline
AT20G/6dF\\
W1             & 61.9\% & 88.7\% & 95.1\%\\
W2             & 67.7\% & 91.1\% & 97.8\%\\
\hline
SDSS-DR12 quasar\\
W1             & 50.8\% & 83.1\% & 95.9\%\\
W2             & 49.3\% & 83.4\% & 96.1\%\\
\hline
\end{tabular}
\end{table}

\begin{figure*}
\begin{minipage}{1.0\textwidth}
\includegraphics[width=1.0\linewidth]{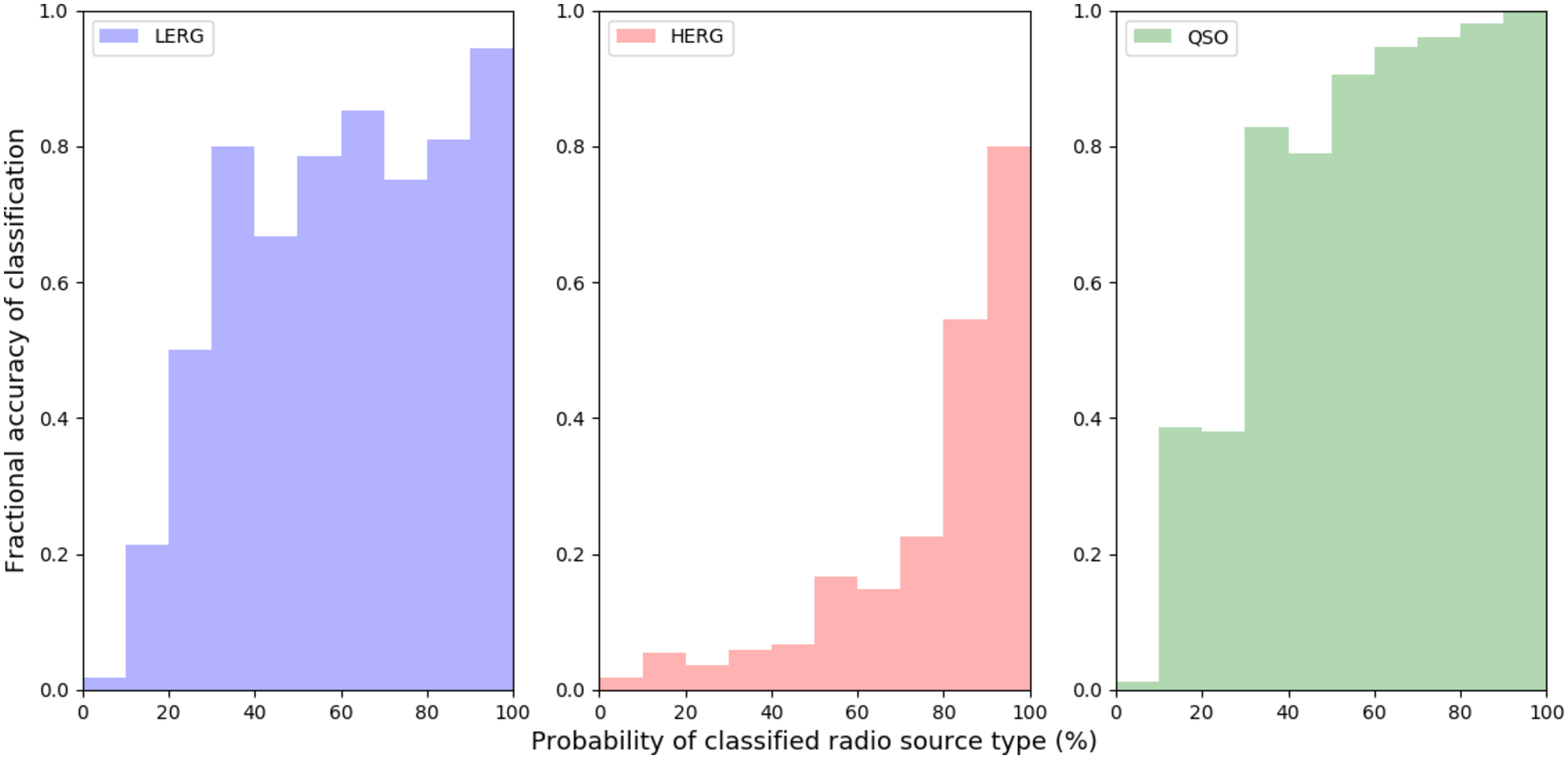}
\caption{Fraction of correct matches of the spectroscopic class identification for the sample selected from  \citet{Jones2009} and \citet{Mahony2011}, with probability of the source being a LERG (left panel), HERG (middle panel) and QSO (right panel). The fraction of successful matches is worst for the HERGs, which is attributed to the overlap of the HERG population in the LARGESS sample with LERGs and QSOs (Fig.~\ref{fig:wisecolours}) and the lower number of HERGs within the training set.}
\label{fig:classcomp}
\end{minipage}
\end{figure*}

In Table~\ref{tab:classrates} we give the success rates for the radio classification by WISE colour for the Best \& Heckman and AT20G/6dF test samples. Overall our classifications made to these samples were accurate in identifying the most likely class. In the latter sample, the results assume uncertain spectral classifications to be correct (for example, objects in the AT20G/6dF sample thought to only have absorption lines in their optical spectra with some uncertainty, indicated by the classification of `Aa?', were assumed to be correct and hence to be a LERG). We also ignored sources with different classifications to those of HERG/LERG/QSO, given the spread in colour space for e.g. star-forming (SF) galaxies. Fig.~\ref{fig:classcomp} shows the accuracy of the match with the probability of the source identification for the AT20G/6dF test sample. Sources spectroscopically identified as HERGs had the worst classification accuracy, attributed to the overlap in WISE colour-colour space with both LERGs and QSOs (see Fig.~\ref{fig:wisecolours}) and the smaller population of HERGs within the LARGESS training set.

For the SDSS-DR12 quasar sample which had no distinction between LERG, HERG and QSO, our program determined that 61.7\% of the sources were most likely to be a QSO from their WISE information. The remainder were most likely HERGs, which is expected given the overlap of these two classes in WISE colour space (Fig.~\ref{fig:wisecolours}) and their similarities. Only four objects of the 2,761 sources were most likely to be a LERG.

\subsection{Redshift estimates}

\begin{table*}
\centering
\caption{The average median and confidence interval range of redshift probability distributions generated for the AT20G/6dF test sample. The redshift probability distributions created from the W2 magnitude tended to be wider and give lower median redshifts than those generated from the W1 magnitude. Overall redshift predictions from W1 and W2 had good agreement. This trend is seen in other test samples (see also Table~\ref{tab:redshiftsucc}).}
\label{tab:redshiftrange}
\begin{tabular}{lllllllll}
\hline
Class & W1 & & & & W2 & & & \\
 & Median & 68\%          & 95\%          & 99.7\%       & Median & 68\%          & 95\%          & 99.7\%       \\
 \hline
All   & 0.35  & 0.16 - 0.70 & 0.06 - 1.09 & 0.01 - 1.55 & 0.35  & 0.13 - 0.71 & 0.04 - 1.12 & 0.00 - 1.68 \\
LERG  & 0.08  & 0.05 - 0.10 & 0.03 - 0.15 & 0.01 - 0.35 & 0.08 & 0.05 - 0.13 & 0.02 - 0.21 & 0.00 - 0.51 \\
HERG  & 0.28  & 0.16 - 0.59 & 0.07 - 1.06 & 0.01 - 1.54 & 0.26  & 0.15 - 0.63 & 0.06 - 1.18 & 0.01 - 1.93 \\
QSO   & 0.65  & 0.25 - 1.31 & 0.07 - 1.95 & 0.00 - 2.64 & 0.64  & 0.20 - 1.29 & 0.05 - 1.91 & 0.00 - 2.62\\
\hline
\end{tabular}
\end{table*}

Table~\ref{tab:redshiftsucc} lists the success rates of the spectroscopically identified redshifts to fall within the 68-95-99.7\% confidence intervals about the median of the redshift probability distributions for W1 and W2. For the SDSS-DR12 QSO sample, the classification of some objects as most likely to be HERGs rather than QSOs does affect the accuracy of our redshift probability distributions within the 68\% confidence interval relative to the other test samples. One consideration is that the QSOs have a far wider redshift space than the HERGs and LERGs in LARGESS (upward of 5 as opposed to 0.8). Even with extrapolation for fainter WISE magnitudes, objects predicted to be a QSO by their WISE colours invariably have wider probability distributions in redshift. However, comparable success rates are seen for the 95\% and 99.97\% confidence intervals. Overall the high agreement between the redshift probability distribution and the spectroscopic redshift gives support to the method we employed. Some discrepancies may be due to differences in the training sets to these tested samples, one which did not offer a LERG or HERG classification to test against (SDSS-DR12 quasar sample).

The use of extrapolation from the well sampled WISE-redshift space of the LARGESS sample (Section 2.3.1) resulted in a greater ability to generate a probability distribution for sources with faint or bright WISE magnitudes. For the AT20G/6dF selected sample, out of 1,960 sources with good quality WISE information, without extrapolation, only 1,710 and 1,706 sources had a probability distribution in redshift created for the W1 and W2 bands respectively ($\sim$87\%), and with extrapolation probability distributions were generated for 1,941 and 1,937 sources ($\sim$99\%). Similar values were seen for the success rates; while they are at worst a few percent lower overall for the 68\% interval, they are on par for the 95\% and better for the 99.7\% intervals. 

While there is the caveat of the trend being assumed to hold for faint or bright WISE magnitudes in our extrapolation, a reasonable approximation can still be obtained for these sources. Sources for which the method failed still had very faint or bright WISE magnitudes, and so these can be assumed to be at either higher or very low redshift respectively.

\subsubsection{Agreement between W1 and W1 bands}

The redshift results from the W1 and W2 bands are consistent with each other. For the AT20G/6dF sample, the average median redshifts for the W1 and W2 bands are both z~=~0.35 (see Table~\ref{tab:redshiftrange} for a full comparison). For the SDSS-DR12 quasar sample, the average median redshifts z~=~0.90 and z~=~0.81, and average 68\% confidence intervals of 0.44~$<$~z~$<$~1.76 and 0.36~$<$~z~$<$~1.73 respectively. 

The redshift probability distributions tend to be slightly wider for those generated from the W2 information than for W1. This resulted in higher likelihoods for the spectroscopically determined redshifts to lie in W2 confidence intervals than for those generated from the W1 magnitude (Table~\ref{tab:redshiftsucc}). Objects found most likely to be LERGs have narrower possible redshift range due to the population within the LARGESS training set, and owing to the redshift relation span for QSOs, objects identified as such have the widest possible redshift ranges.

\subsection{Test with CENSORS}

\begin{figure}
\includegraphics[width=1.0\linewidth]{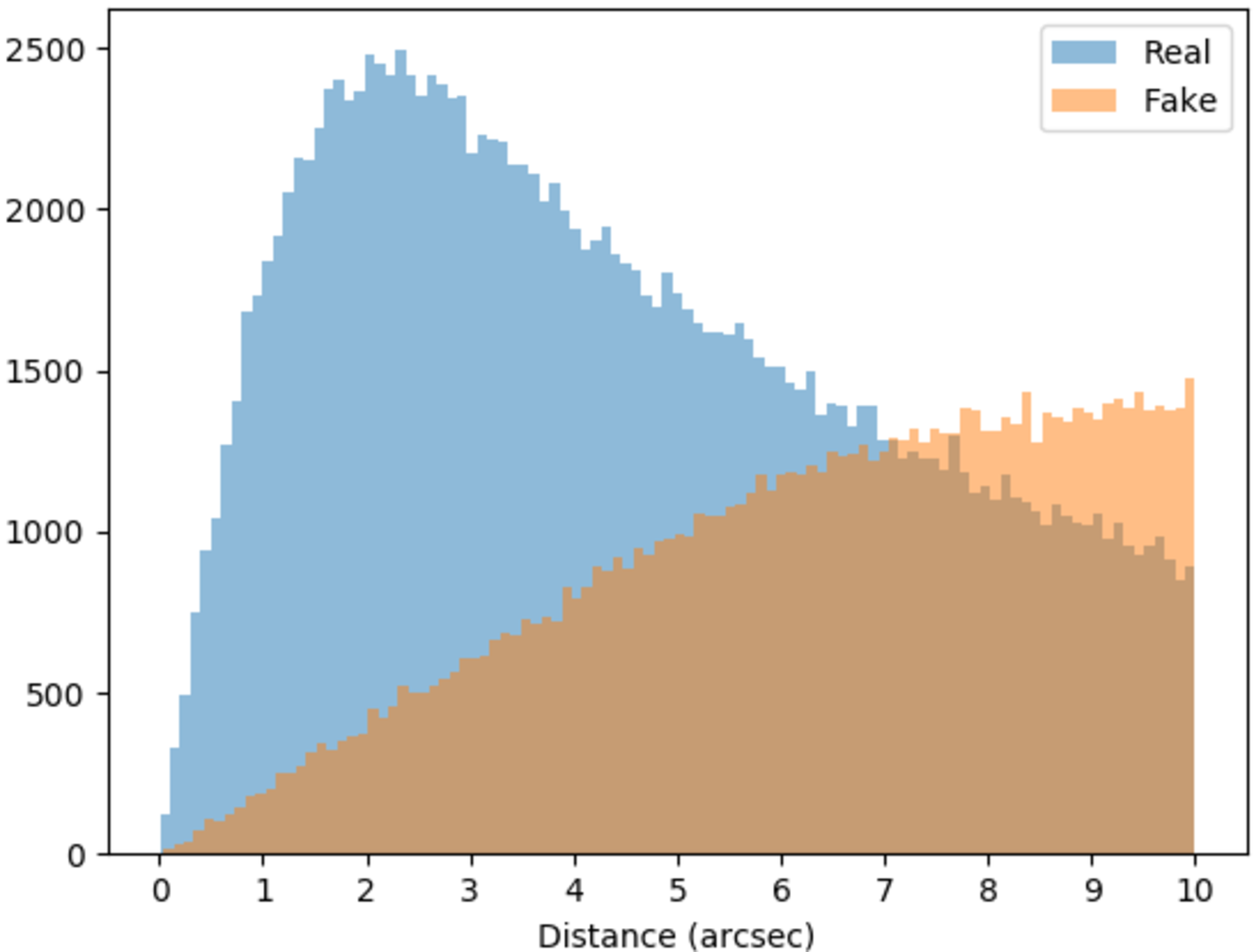}
\caption{Histograms of the separation between input positions and a WISE match for sources within the SUMSS survey (real) and randomly generated fake positions. An overlap of the two histograms at 7 arcseconds indicates at higher search radii you are less likely to find a real WISE association with a radio source.}
\label{fig:separation}
\end{figure}

To compare with existing deep surveys with known spectroscopic redshifts and examine whether our redshift estimation method can reproduce observations of radio galaxies, we cross-matched the SUMSS catalogue with WISE. We first considered at what radius to conduct any cross-match with WISE. Fig.~\ref{fig:separation} shows the number of detections with separation for both the SUMSS sample and a `fake' sample constructed with random positions (uniformly distributed). The crossover at 7\,arcsec suggests that you are less likely to find a real association with a radio source at higher separation distances. We opted for a crossmatch upper limit of 2.5\,arcsec, to ensure a higher reliability (87\% at 2.5\,arcsec compared to 72\% at 7.0\,arcsec) at the cost of completeness (28\% versus 79\%). Like with the LARGESS sample, we also required a signal-to-noise ratio of at least two in the W1, W2 and W3 WISE bands. In total we had 15,043 sources.

\begin{figure*}
\includegraphics[width=1.0\linewidth]{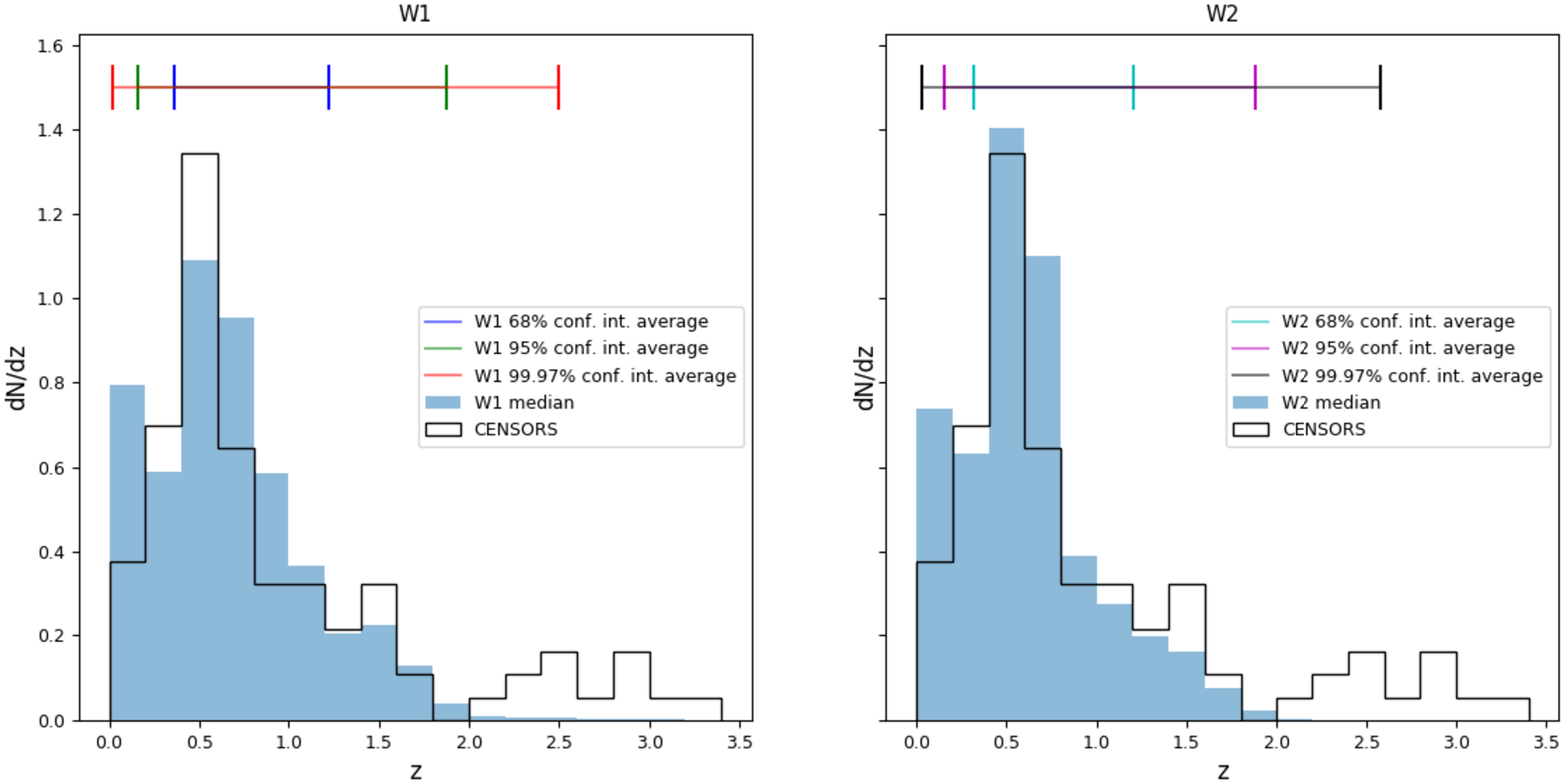}
\caption{Comparison of the median redshift of each probability redshift distribution for SUMSS sources with a flux density $S_{\rm{843 MHz}}~>$~10~mJy with WISE data for both the W1 and W2 bands (left and right panels respectively), with redshifts from the CENSORS survey \citep[black unfilled histogram;][]{Brookes2008}. There is reasonable agreement up to z~$\sim$~2, where the median fails to replicate the high redshift population found in CENSORS. However, the average of the 95\% and 99.97\% confidence intervals for our redshift probability distributions extend beyond z = 2. SUMSS sources with unreliable cross-matches and/or poor quality WISE measurements are not included; such sources are infrared faint and likely to also lie at higher redshifts.}
\label{fig:censors}
\end{figure*}

We compare our redshift probability distributions for SUMSS with that of the Combined EIS-NVSS Survey Of Radio Sources \cite[CENSORS;][]{Brookes2008}, a 1.4~GHz radio survey for a 6-degree field to a flux density of 7.2~mJy. In fig.~11 of \cite{deZotti2010} a redshift distribution for CENSORS sources brighter than 10~mJy are compared to redshift distribution models for radio sources. Fig.~\ref{fig:censors} gives the distribution of median redshifts for each probability redshift density function for sources with a flux density of 10~mJy or brighter in SUMSS (0.843~GHz).

Up to redshift of z~$\sim$~2 there is good agreement with our distributions and that of CENSORS sources brighter than 10~mJy at 1.4 GHz. We conduct the the Kolmogorov-Smirnov test between the two distributions, which is a non-parametric test comparing the distance between the empirical distribution functions of two samples. For z~$<$~2, we obtain a K-S statistic of 0.098, or p-value of 0.4, confirming our method matches well with the CENSORS survey up to this redshift limit. 

It is noted that our distributions include a peak at z~=~0 while CENSORS does not. It is possible that the LARGESS sample has a higher population of low-redshift radio sources, or that we are biasing towards LERGs in radio classification. Alternatively, a number of sources are neither HERG, LERG or QSO but rather a local star-forming population we cannot distinguish through WISE colours from the other populations (as identified in fig. 15 of \cite{Ching2017}).

Few sources are predicted through our code to have a median redshift greater than z~=~2.0 like that seen in the CENSORS survey, but the average of the 95\% and 99.7\% confidence intervals extends to higher redshifts. When considering the full distributions the K-S two sample statistic is 0.157, giving a corresponding p-value of 0.02. It should be noted that the median redshift is merely an indicator and not the most likely redshift value. Another consideration is that we filtered out unreliable WISE magnitudes; that is, filtered out lower signal-to-noise sources which often have fainter or upper limits in their magnitudes, which by the WISE magnitude-z relation would suggest higher redshift sources being filtered out \cite[e.g. infrared-faint radio sources;][]{Collier2014}. The cross-match rate between CENSORS and WISE also decreases with redshift in line with our predictions for SUMSS sources (83\% with a WISE cross-match within 2.5\,arcsec for CENSORS sources with z~$<$~1, 69\% for 1~$<$~z~$<$~2, and 13\% for z~$>$~2 - see Fig.~\ref{fig:complete}).

\subsection{Comparison with other possible methods}\label{othermethods}

We tested other methods for both the radio class classification and the redshift probability estimation. 

\subsubsection{Classification using machine learning}
In order to assess the reliability of our adopted classification method described in Section \ref{kdeclass}, we compare the classification outcomes with an alternative approach using machine learning. Specifically, we adopt the linear discriminant analysis (LDA) algorithm as implemented in {\sc scikit-learn} \citep{Pedregosa2011}. This method uses supervised classification on a training set \cite[in this case][]{Ching2017} to find axes which maximise the separation between any number of given classes across $N$-dimensions, the results of which can then be applied to new data to offer probabilistic classification of each class.

\begin{table}
\centering
\caption{Radio classification success rates for the blind tests of the Best \& Heckman (2012) sample and the AT20G/6dF sample, based on using the machine learning approach described in Section \ref{othermethods}. Compare to Table~\ref{tab:classrates} in which the classifications were done through KDE. While the LERG and QSO identification success rates are consistently slightly  higher, the HERG identification success rate is significantly lower for the Best \& Heckman sample.}
\label{tab:class-ml}
\begin{tabular}{lrrrr}
\hline
Sample & LERG & HERG & QSO & Total\\
\hline
Best \& Heckman sample & 97.6\% & 30.0\% & - & 94.1\%\\
AT20G/6dF sample & 97.8\% & 50.9\% & 90.4\% & 90.7\%\\
\hline
\end{tabular}
\end{table}
\begin{table}
\centering
\caption{ Redshift estimation success rates for the  three test samples following radio classification via the LDA method. Compare to Table~\ref{tab:redshiftsucc} in which the classifications were done through KDE. Similar success rates are seen throughout.}
\label{tab:redshiftsucc-ml}
\begin{tabular}{lrrr}
\hline
WISE magnitude & 68\% & 95\% & 99.7\%\\
\hline
Best \& Heckman\\
W1             & 71.8\% & 96.7\% & 99.8\%\\
W2             & 78.3\% & 99.0\% & 99.9\%\\
\hline
AT20G/6dF\\
W1             & 64.4\% & 91.4\% & 98.5\%\\
W2             & 70.2\% & 93.4\% & 98.8\%\\
\hline
SDSS-DR12 quasar\\
W1             & 55.6\% & 84.8\% & 95.1\%\\
W2             & 54.8\% & 85.2\% & 95.7\%\\
\hline
\end{tabular}
\end{table}

In our definition of the LDA classification, we use the W1/W2/W3 magnitudes, as well as the W1-W2 and W2-W3 colours. The classification success rates through LDA are provided in Table~\ref{tab:class-ml}, while those made through the kernel density estimation (KDE) method are in Table~\ref{tab:classrates}. When applied to the AT20G/6dF sample, an overall classification success rate of 91\% was obtained for the spectroscopically identified radio galaxies (versus 87\% through KDE). Minor improvements were made for the LERG and QSO objects, and equal rates found for the HERGs. For the Best \& Heckman sample, the overall classification success rate through LDA was 94\% (versus 89\%), but significantly worse for the HERG population (30\% versus 61\%). This sample had only a low number of HERGs, so the overall success rate reflects the higher success rate in classifying the LERG population. Lastly, for the SDSS-DR12 quasar sample, the LDA method predicted 200 sources to be most likely LERGs (compared to the KDE method employed on the WISE colours which predicted only 4 objects to be LERGs), and more QSOs (2,412, or 87\% compared to only a radio QSO population of 62\% predicted through KDE).

As evident from the Best \& Heckman sample results, the machine learning approach preferentially picked sources to be LERGs rather than HERGs. The average likelihood for any of the sources in the AT20G/6dF sample to be a LERG was $\sim$43\% with the LDA method, compared to $\sim$37\% for the KDE method; for HERGs the average likelihood was $\sim$24\% versus $\sim$29\%. The discrepancy in the HERGs may be partly due to the overlap in colour space with the other classes, and in addition the similarity in W1 magnitudes between HERGs and LERGs within LARGESS. When considering the mid-infrared colour information alone, we found the LDA approach was correspondingly less likely to classify sources as HERGs. In comparison, the KDE method is only applied to WISE colour space without incorporating magnitude information, and, by considering the distance in colour space to the peak of the HERG distribution, appears to better take into account the overlap between the three populations. 

Redshift probability distributions derived from both methods were ultimately found to have good agreement. Similar results were found in all the samples for the confidence intervals tested (Table~\ref{tab:redshiftsucc-ml}), even when using both the LDA and KDE methods together. The exception is that higher success rates were found for the 68\% and 95\% confidence intervals for the SDSS-DR12 quasar sample of a few percent. This can likely be attributed to wider average ranges of redshift for the probability density functions using the machine learning method. For example, the 68\% confidence interval for the KDE method gave an average redshift range of z$_{range}$~=~1.31 and 1.37 in W1 and W2 respectively, and z$_{range}$~=~1.41 and 1.50 for the LDA approach. This means the redshift probability distributions generated from LDA classifications are wider. The increase in redshift range from the LDA approach is hence possibly due to the higher identification rate of objects to be QSOs through that method than LERGs or HERGs, which span a larger redshift range in our training sample from LARGESS (Fig.~\ref{fig:WISE_z_fit}).  

The general agreement of the redshift estimates between the two methods indicates that classifying the radio galaxies prior to making a redshift estimation is a sound approach to take, and that both the KDE method and LDA machine learning approach offer similar accuracies in the initial classification stage.

\subsubsection{kth nearest neighbour}

We also considered a kth nearest neighbour approach for redshift estimation, where we took the redshifts of ten LARGESS objects for each class closest in W1 and W2 magnitude to an input source and calculated the mean and standard deviation. While this method still produces a redshift estimate for faint and bright W1 or W2 magnitudes, these estimates are made for the extreme sources in the sample only which are limited in redshift, and hence will suffer in accuracy.

The kth nearest neighbour method we considered for the test sample constructed from AT20G/6dF only had 65.3\% and 58.8\% success rates (for W1 and W2 respectively) for the spectroscopic redshift to fall within the standard deviation of the mean of the 10th nearest neighbour. While this is not far from the values we calculated for the 68\% confidence intervals (Table~\ref{tab:redshiftsucc}), it fails to account for the full possible redshift space for the source. This is only provided by constructing a redshift probability distribution which already weighs by the chance of each radio classification. Furthermore, the kth nearest neighbour method is less accurate in estimating the redshift for sources with very faint or bright W1 magnitudes, where there is a cut-off in sources within the LARGESS sample to use in the estimate without any use of extrapolation, which we adopted for our chosen method.

\section{Predictions}

Here we now consider two potential applications of our code to make predictions. The first explores the percentage of radio sources we can expect to see within the redshift range where associated H{\sc i} absorption can be detected by the FLASH survey, and what class of radio AGN these fall into. In the second case, we examine whether low-frequency peaked spectrum sources, potentially young or recently re-triggered radio AGN, are likely to be found at high redshift. 

\subsection{Predictions for FLASH}

Surveys for neutral hydrogen (H{\sc i}) absorption can inform us on the evolution of the cold gas with redshift that fuels the accretion of AGN, as well as how it is impacted by AGN feedback. Examples of large upcoming H{\sc i} absorption surveys that will address this topic, where H{\sc i} gas is detected towards sufficiently bright background radio sources through the 21~cm transition, include the First Large Absorption Survey in H{\sc i} (FLASH; Sadler et al.) and Widefield ASKAP L-Band Legacy All-Sky Blind Survey \cite[WALLABY;][]{Koribalski2012} with the Australian Square Kilometre Array Pathfinder \cite[ASKAP;][]{Deboer2009, Johnston2009, Schinckel2012}, the Search for H{\sc i} absorption with APERTIF (SHARP; Morganti et al.), and the MeerKAT Absorption Line Survey \cite[MALS;][]{Gupta2017}.

FLASH will search within the redshift space of 0.4~$<$~z~$<$~1 survey for both associated and intervening absorption toward radio-bright objects (flux density of $\sim$~50 mJy and greater). However, as most of these objects lack optical identifications or reliable redshift information, any H{\sc i} detection made requires follow-up to verify whether it is associated with the host radio galaxy or merely within a galaxy along the line of sight. Optical follow-up has already been required (and undertaken) for PKS~1740-517, the first new HI absorption detection made with ASKAP during commissioning \citep{Allison2015}. Knowledge of the redshift distribution of the background radio sources is vital to understanding the average H{\sc i} spin temperature \citep{Allison2016}, and can statistically determine the redshift interval probed by the radio source population. This complements efforts to distinguish associated and intervening absorption systems through machine learning algorithms \citep{Curran2016}.

Using the cross-matched SUMSS and WISE sample (Section~3.4 - 15,043 radio sources), 19.3\% were found to likely be a LERG, 49.0\% a HERG and 31.0\% a QSO, with the remaining 0.7\% (107 sources) unable to be reliably identified by their WISE colour information. The average median redshifts found from W1 and W2 were z = 0.66 and 0.60 respectively. Note that these are only SUMSS sources with a reliable WISE counterpart.

Table~\ref{tab:sumssprob} gives the probability for sources within the sample to have a redshift at z~$<$~0.4, 0.4~$<$~z~$<$~1.0, and z~$>$~1.0. In total there is a 46.1\% chance for a radio source in SUMSS with a WISE counterpart to lie within the redshift range which can be searched for associated H{\sc i} absorption in FLASH. Therefore, roughly 46\% of radio sources with a reliable cross-match with WISE ($\sim$7,000) will be able to be searched for H{\sc i} associated absorption by FLASH, although realistically only those with a minimum flux density of around 50~mJy will be bright enough to sufficiently search towards (that is, $\sim$3,200 radio sources with reliable and good-quality WISE information that can be searched for associated H{\sc i} absorption).

\begin{table}
\centering
\caption{The average probability for a radio source \textit{with a WISE counterpart within 2.5\,arcsec} in SUMSS to have a redshift either lower than, higher than, or within the redshift range for the FLASH project with ASKAP. Infrared-faint radio sources are more likely to be at higher redshifts \citep[Fig.~\ref{fig:WISE_z_fit} and e.g.][]{Collier2014}. Those at z $<$ 0.4 will be unable to be searched for associated H{\sc i} absorption within the FLASH project. These values can be calculated for other redshift values via our code, as desired set by the user.}
\label{tab:sumssprob}
\begin{tabular}{llrrr}
\hline
Minimum Flux & Class & z $<$ 0.4 & 0.4 $<$ z $<$ 1.0 & z $>$ 1.0 \\
\hline
No flux cut & All & 35.5\% & 46.1\% & 18.4\% \\
 & LERG & 72.3\% & 27.5\% & 0.2\% \\
 & HERG & 28.3\% & 60.6\% & 11.1\% \\
 & QSO & 15.4\% & 36.0\% & 48.6\% \\
 \hline
$\geqslant$20 mJy & All & 33.1\% & 46.1\% & 20.8\% \\
 & LERG & 74.1\% & 25.7\% & 0.2\% \\
 & HERG & 25.5\% & 62.1\% & 12.4\% \\
 & QSO & 15.7\% & 36.2\% & 48.1\% \\
 \hline
$\geqslant$50 mJy & All & 31.4\% & 45.8\% & 22.8\% \\
 & LERG & 76.2\% & 23.6\% & 0.2\% \\
 & HERG & 24.7\% & 62.0\% & 13.3\% \\
 & QSO & 17.1\% & 36.6\% & 46.3\%\\
 \hline
\end{tabular}
\end{table}

This percentage varies both for the class of object predicted by the WISE colour information for the source; objects identified to be a QSO have a higher chance of a greater redshift value, as expected due to the trends seen in the LARGESS training sample. A cut to higher flux densities slightly shifts sources to higher redshifts (we calculate a 22.8\% probability of sources both likely to be a QSO and a flux density greater than 50 mJy to like at a redshift beyond z~=~1, as opposed to 18.4\% for QSOs without that flux cut).

We also see a lower percentage of LERGs and higher percentage of HERGs in the 0.4~$<$~z~$<$~1 range when comparing the 50~mJy flux density cut to all sources. This agrees with the fall-off in LERG populations with redshift and the luminosity evolution of HERGs with redshift \cite[figs. 2 and 10 respectively of][]{Pracy2016}. This also supports the differences found between the LERG and HERG populations \cite[see also][]{Best2012}.

It should be noted, however, that like LARGESS, the number of LERGs diminished at similar redshift limits due to sensitivity limits of the \cite{Pracy2016} study. We also may be underestimating the number of HERGs (see Table~\ref{tab:redshiftsucc}). Therefore the LARGESS training set may be biasing our predictions here. The effect in Fig.~\ref{fig:complete}, where a slight decrease in the fraction of LARGESS sources with WISE cross-matches made is seen with increasing redshift, should also be considered.

\subsection{Sample of low-frequency peaked-spectrum radio sources}

Of particular interest to understanding AGN evolution is the population of high redshift radio galaxies \citep{Miley2008}. Radio sources with a peak in their spectral energy distribution (e.g. Gigahertz-peaked spectrum sources; GPS) are believed to be young or recently re-triggered radio AGN \cite[e.g.][]{ODea1991}. Sources peaked at lower frequencies (Megahertz-peaked spectrum sources; MPS) have been proposed to be these young radio AGN at high redshift (z~$>$~2) such that the turnover frequency has shifted to this lower regime \citep{Falcke2004,Coppejans2015,Coppejans2016,Callingham2017}. Therefore, by identifying the redshifts for these galaxies, we can better learn more about the evolution of radio AGN with redshift. We examine the WISE properties of such low-frequency peaked spectrum sources to see whether their mid-infrared information supports the hypothesis that they lie at high redshift.

85 low-frequency peaked-spectrum radio sources with a match to NVSS or SUMSS were selected from the study of \cite{Callingham2017}. Of this sample, only 48 sources had a match with WISE within 10\,arcsec of their NVSS or SUMSS position, and 26 within 2.5\,arcsec. The majority of these had poor quality WISE magnitudes, with upper limits in W3 or even the W2 magnitudes. In total only 9 sources were found with a WISE cross-match within 2.5\,arcsec and with sufficient quality WISE magnitude information (13 within 10\,arcsec, or 39 sources within 10\,arcsec with poor quality WISE data). 

Of the 9 sources we do have sufficient WISE information for, 2 have over 10\% of their probability density functions supporting the hypothesis that the source is at z~$>$~2 (the higher with 26-30\% probability). 14 of the other 39 sources successfully cross-matched with WISE at higher cross-match radii and/or with upper limits for their W2 or W3 magnitudes were still found with a significant probability ($>$~10\%) to lie at z~$>$~2. We note again these are either unreliable cross-matches and hence of these 39 sources, many may be simply undetected in WISE, or redshift estimates for sources with upper limits in their W2 magnitudes are hence biased to lower values. All remaining sources with no cross-match made are faint in WISE and hence likely lie at higher redshift (see Fig.~\ref{fig:WISE_z_fit} and also the study of infrared-faint radio sources by \cite{Collier2014}). Therefore, the majority of the sample are indeed likely to lie at high redshift.

\section{Summary}

We have presented a new method for estimating the redshift for radio galaxies, using the LARGESS sample as a training set and the WISE mid-infrared magnitude information. This method assigns a probability to the LERG, HERG and QSO classes to the radio galaxies, and then generates a redshift probability distribution. The code which generates these probability distributions is publicly available to use\footnote{\url{https://github.com/marcinglowacki/wise_redshift_estimator}}. Best fit equations are also offered as an alternative (Tables~\ref{tab:bestfit} and \ref{tab:loglinfits}).

Testing on samples with known spectroscopic redshift information, and comparison with the CENSORS survey, suggests the probability distributions our algorithms generate do well in constricting the possible redshift for radio galaxies. We predict that most radio sources in SUMSS with a WISE counterpart lie at z~$<$~1, with sources at z~$<$~0.4 dominated by LERGs and z~$>$~1 by QSOs. We predict 46\% of the radio sources with good WISE measurements will be able to be searched for associated H{\sc i} absorption through the FLASH survey, which will aid our understanding of the evolution of the cold star-forming fuel in radio galaxies. We also find most sources found to peak at low frequency in their spectral energy distributions have faint or no WISE cross-matches with their radio positions, and hence are likely to lie at higher redshifts (z~$>$~2). 

This method is ideal for radio galaxies lacking in any optical spectroscopy, which is an area of concern for current and upcoming radio surveys in the southern sky. Further exploration through this method, alongside other photometric redshift estimating techniques, is necessary to narrow down the redshift distribution of such galaxies. However, given the uncertainties associated with photometric estimates (including ours) and the need for details optical information about physical hosts of radio AGN, a focus on obtaining reliable optical spectroscopic redshift information for these galaxies is of vital importance in future radio surveys. 

\section*{Acknowledgements}

We thank John Ching, Scott Croom, Elizabeth Mahony and Jack Singal for useful discussions. We also thank Joseph Callingham for making data available from \cite{Callingham2017} for this research. Parts of this research were conducted by the Australian Research Council Centres of Excellence for All-sky Astrophysics (CAASTRO) and All-sky Astrophysics in 3D (ASTRO 3D), through project numbers CE170100013 and CE110001020. This research has also made use of NASA's Astrophysics Data System Bibliographic Services. This research made use of Astropy, a community-developed core Python package for Astronomy (Astropy Collaboration, 2013).

\footnotesize{
  \bibliographystyle{mn2e}
  \bibliography{bibliography}
}
\label{lastpage}

\end{document}